\documentclass[12pt]{article}
\usepackage{}
\usepackage{geometry}             
\usepackage{graphicx}
\usepackage{amssymb}
\usepackage{amsmath}
\usepackage{epstopdf}
\usepackage{comment}
\usepackage{cite}

\usepackage[hyperindex=true,
          pdfstartview=FitH,
          bookmarksnumbered=true,
          bookmarksopen=true,
          citecolor=blue,
          linkcolor=blue,
          colorlinks=true,
          unicode]{hyperref}

\usepackage{booktabs}
\usepackage[font=small,labelfont=bf]{caption}
\usepackage{makecell,multirow}

\setcellgapes{0.5ex}

\allowdisplaybreaks
\numberwithin{equation}{section}

\begin{document}
\title{$\lambda$ phase transition in Horava gravity}
\author{Wei Xu$$\thanks{{\em
        email}: \href{mailto:xuwei@cug.edu.cn}
        {xuwei@cug.edu.cn}}\\
         School of Mathematics and Physics,\\
China University of Geosciences, Wuhan 430074, China}
\date{}
\maketitle
\begin{abstract}
In this paper, we present another example of superfluid black hole containing $\lambda$ phase transition in Horava gravity. After studying the extended thermodynamics of general dimensional Horava-Lifshitz AdS black holes, it is found that only the one with spherical horizon in four and five dimensions have a $\lambda$ phase transition,
which is a line of (continuous) second order phase transitions
and was famous in the discussion of superfluidity of liquid ${}^4$He. The ``superfluid" black hole phase and  ``normal" black hole phase are also distinguished. Especially, six dimensional Horava-Lifshitz AdS black holes
exhibit infinitely many critical points in $P-\nu$ plane and the divergent points for specific heat,
for which they only contain the ``normal" black hole phase and the ``superfluid" black hole phase disappears due to the physical temperature constraint;
therefore there is no similar phase transition.
In more than six dimensions, there is no $P-\nu$ critical behavior.
After choosing the appropriate ordering field, we study the critical phenomena in different planes of thermodynamical phase space. We also calculate the critical exponents, which are the same with the van der Waals fluid.


\end{abstract}

\section{Introduction}
Black hole thermodynamics always provides valuable insight into quantum properties of gravity,
and it has been studied extensively for quite a long time, especially for the quantum and microscopic interpretation of black hole temperature and entropy (See \cite{Parikh:1999mf,Strominger:1996sh} for example). Besides, thermodynamics and  phase transitions of AdS black holes
have been of great interest since the Hawking-Page phase transition \cite{Hawking:1982dh} between stable large black hole and thermal gas is  explained  as  the  confinement/deconfinement phase transition of gauge field \cite{Witten:1998zw} inspired by the
AdS/CFT correspondence \cite{Maldacena:1997re,Gubser:1998bc,Witten:1998qj}.

After treating the cosmological constant as a pressure with its conjugate quantity being the
thermodynamic volume in thermodynamic phase space of charged AdS black holes \cite{Kastor:2009wy,Dolan:2011xt,Cvetic:2010jb,Dolan:2013ft,Kastor:2010gq,Castro:2013pqa,El-Menoufi:2013pza},  the small/large black hole phase transition is established in \cite{Kubiznak:2012wp}, which is exactly
analogy with the liquid/gas phase transition of the van der Waals fluid.
This kind of black hole phase transitions have attracted much attention (See the recent review papers \cite{Altamirano:2014tva,Kubiznak:2016qmn}).
Besides, this semiclassical method of analogue is generalized to
study the microscopic structure of black holes, and sheds some light on the
black hole molecules \cite{Wei:2015iwa} and microscopic origin
of the black hole reentrant phase transition \cite{KordZangeneh:2017lgs}.
The study is also extended to the quantum statistic viewpoint, as
the superfluid black holes are reported recently \cite{Hennigar:2016xwd}.
In Lovelock gravity with conformally coupled scalar field, the authors present the first example of a $\lambda$  phase transition, which is a line of (continuous) second order phase transitions
and was famous for the successful quantum and microscopic interpretation of superfluidity of liquid ${}^4$He.

In this paper, we present another example of black holes containing $\lambda$ phase transition in Horava gravity, which is a candidate of quantum gravity in ultra high energy \cite{Horava:2009uw}. The Horava-Lifshitz (HL) black hole solutions, thermodynamics and phase transitions have attracted a lot of attention \cite{Lu:2009em,Cai:2009pe,Cai:2009qs,Myung:2009va,Cao:2010ft,Majhi:2012fz,Mo:2013sxa,Ma:2017pap} (See \cite{Wang:2017brl} for a review on the recent development of various areas). The general dimensional HL black hole solutions are also introduced \cite{Li:2014fsa}.
We will consider the extended thermodynamics of general dimensional HL AdS black holes. It is shown that only the one with spherical horizon in four and five dimensions have a $\lambda$ phase transition. Note that the first example of ``superfluid" black holes always have a hyperbolic horizon\cite{Hennigar:2016xwd}.
Especially, six dimensional HL AdS black holes
exhibit infinitely many critical points in $P-\nu$ plane and the divergent points for specific heat, for which they only contain the ``normal" black hole phase and the ``superfluid" black hole phase disappears due to the physical temperature constraint;
therefore there is no similar phase transition.
In more than six dimensions, there is no $P-\nu$ critical behavior.
After identifying parameter $\epsilon$ as the ordering field instead of pressure and temperature, we study the critical phenomena in different planes of thermodynamical phase space. We also obtain the critical exponents, which are the same as the van der Waals fluid.

The paper is structures as follows: in next Section, we present the extended thermodynamics of generalized topological HL black holes; Then we study $P-V$ criticality in Section 3;
We show the $\lambda$ phase transition for four and five dimensions and the discussion for six dimensions in Section 4 and Section 5, respectively; In Section 6 and Section 7, we discuss the critical phenomena and calculate the critical exponents in different planes; In final Section, some concluding remarks are given.

\section{Extended thermodynamics of generalized HL black holes}
In this section, we present the extended thermodynamics of generalized topological HL black holes in $(d+1)$ dimensions ($d\geq3$). We begin with the action of HL gravity at the $z=3$ UV fixed point, which can be re-expressed as \cite{Li:2014fsa}
\begin{align}
  S&=\int\mathrm{d}t\bigg[L_0+(1-\epsilon^2)L_1\bigg],\label{action}\\
  L_0&\equiv\int\mathrm{d}^{d}x\sqrt{g}N\bigg[\frac{2}{\kappa^2}\bigg(K_{ij}K^{ij}-\lambda\,K^2\bigg)
  +\frac{\kappa^2}{8\kappa^4_{W}}\frac{\Lambda_{W}}{1-d\lambda}\bigg((d-2)R-d\Lambda_{W}\bigg) \bigg],\nonumber\\
  L_1&\equiv\int\mathrm{d}^{d}x\sqrt{g}N\frac{\kappa^2}{8\kappa^4_{W}}\frac{1}{1-d\lambda}
  \bigg[\left(1-\frac{d}{4}-\lambda\right)R^2-(1-d\lambda)R^{ij}R_{ij} \bigg],\nonumber
\end{align}
where the first two terms in the $L_0$ are the kinetic actions, while
the residue correspond to the potential actions. $R_{ij}$ is the Ricci tensor, $R$ is the Ricci scalar and $K_{ij}$
is defined by $K_{ij}=\frac{1}{2N}\bigg(\dot{g}_{ij}-\nabla_{i}N_{j}-\nabla_{j}N_{i}\bigg)$, which are based on the ADM decomposition of the higher dimensional metric, i.e. $\mathrm{d}s^2_{d+1}=-N^2\mathrm{d}t^2+g_{ij}(\mathrm{d}x^i-N^i\mathrm{d}t)(\mathrm{d}x^j-N^j\mathrm{d}t)$.
Here the lapse, shift and $d$-metric $N$, $N^i$ and $g_{ij}$ are all functions of $t$ and $x^i$, and a dot denotes a derivative with respect to $t$. There are five constant parameters in the action: $\Lambda_{W}$, $\lambda$, $\epsilon$ $\kappa$ and $\kappa_{W}$. $\Lambda=\frac{d}{2(d-2)}\Lambda_{W}$ is the cosmological constant. $\kappa$ and $\kappa_{W}$ have their origin as the Newton constant and  the speed of light.
 $\lambda$ represents a dynamical coupling constant which is susceptible to quantum corrections.
We will fix $\lambda=1$ in the following paper, only for which general relativity can be recovered in the large distance approximation.
In addition, we will only consider the general values of $\epsilon$ in the region $0\leq\epsilon^2\leq1$,
as $\epsilon=0$ corresponds to the so-called detailed-balance condition, and HL gravity with $\epsilon=1$ returns back to general relativity.

The action Eq.(\ref{action}) admits a arbitrary dimensional topological AdS black holes with the metric
\begin{align}
  \mathrm{d}s^2=-f(r)\mathrm{d}t^2+\frac{\mathrm{d}r^2}{f(r)}+r^2\mathrm{d}\Omega^2_{d-1,k}
\end{align}
and the horizon function \cite{Cai:2009pe,Li:2014fsa}
\begin{align*}
  f(r)=k-\frac{2\Lambda_{W}}{(1-\epsilon^2)}\frac{r^2}{(d-1)(d-2)}
  -r^{2-\frac{d}{2}}\sqrt{\frac{c_0}{(1-\epsilon^2)}+\frac{\epsilon^2}{(1-\epsilon^2)^2}\frac{4\Lambda_{W}^2r^{d}}{(d-1)^2(d-2)^2}},
\end{align*}
where $\mathrm{d}\Omega^2_{d-1,k}$ denotes the line element of a $(d-1)$
dimensional manifold with constant scalar curvature $(d-1)k$,
and $k=0, \pm1$ indicates different topology of the spatial space.
$c_0$ is integration constant, which is related to the black hole mass
\begin{align}
  M=-\frac{\Omega^2_{d-1,k}c^3}{16\pi\,G_{N}}\frac{1}{(d-2)\Lambda_{W}}c_0=-\frac{c_0}{\Lambda_{W}},
\end{align}
where we have chosen the natural units and $\Omega^2_{d-1,k}=16(d-2)\pi$.

In AdS spacetime, the cosmological constant is introduced as the thermodynamical pressure \cite{Kubiznak:2012wp}
\begin{align}
  P=-\frac{\Lambda}{8\pi}=-\frac{d}{16\pi(d-2)}\Lambda_{W}.
\end{align}
Then we can re-write the horizon function as
\begin{align}
  f(r)=k+\frac{32\pi\,Pr^2}{(1-\epsilon^2)d(d-1)}
  -4r^{2-\frac{d}{2}}\sqrt{\frac{(d-2)MP\pi}{d(1-\epsilon^2)}+\frac{64\epsilon^2P^2\pi^2r^{d}}{(1-\epsilon^2)^2d^2(d-1)^2}},
\end{align}
and the mass is
\begin{align}
M=\frac{64P\pi\,r_{+}^d}{(d-1)^2(d-2)d}+\frac{(1-\epsilon^2)dk^2r_{+}^{d-4}}{16P(d-2)\pi}+\frac{4kr_{+}^{d-2}}{(d-2)(d-2)},
\end{align}
where $r_{+}$ denotes the event horizon which is the largest positive root of $f(r)=0$.
The conjugate thermodynamic volume of pressure is
\begin{align}
  V=\frac{64\pi\,r_{+}^{d}}{(d-1)^2(d-2)d}-\frac{(1-\epsilon^2)dk^2r_{+}^{d-4}}{P^2(d-2)\pi}.
\end{align}
The entropy and temperature are presented in \cite{Li:2014fsa} with the following forms:
\begin{equation}
S=
\left\{ \begin{array}{ll}
4\pi\,r_+^2\bigg(1+\frac{kd(1-\epsilon^2)\ln(r_+)}{8(d-2)P\pi\,r_+^2}\bigg)+S_0,
\qquad\qquad\quad d=3, \\
\frac{16\pi\,r_+^{d-1}}{(d-1)^2(d-2)}\bigg(1+\frac{kd(d-1)^2(d-2)(1-\epsilon^2)}{32(d-2)(d-3)P\pi\,r_+^2}\bigg)+S_0, \quad d\geq4
\end{array} \right.
\end{equation}
and
\begin{align}\label{temperature}
  T=\frac{1024P^2\pi^2r_+^4+64k(d-1)(d-2)P\pi\,r_+^2+k^2d(d-1)^2(d-4)(1-\epsilon^2)}{8(d-1)\pi\,r_+\bigg(32\pi\,Pr_+^2+kd(d-1)(1-\epsilon^2)\bigg)}.
\end{align}
It is easy to check the first law of thermodynamics
\begin{align}
  \mathrm{d} M=T\mathrm{d}S+V\mathrm{d}P,
\end{align}
while the Smarr relation always fails to exist. This can be easily found
because of the existence of $S_0$ (and logarithmic term for $d=3$) in black hole entropy,
which is not fixed and may be calculated by invoking the quantum theory of gravity as argued in \cite{Cai:2009pe}.
Note that $\epsilon$ scales as $[L]^{0}$, there is no other dimensional quantity in
extended thermodynamic phase space. In this meaning, the validity of the
Smarr relation will bring another physical consideration on the parameter $S_0$.

In order to analyze the global thermodynamic stability and phase transition of the HL black
hole, it is always to study the Gibbs free energy
\begin{align}
  G=H-TS=M-TS,
\end{align}
as the black hole mass $M$ should be considered as the enthalpy $H$ in the extended thermodynamic phase space.
For the local thermodynamic stability, one can turn to the specific heat of black hole
\begin{align}
  C=\frac{\mathrm{d}M}{\mathrm{d}T}=\frac{\mathrm{d}M}{\mathrm{d}r_{+}}/\frac{\mathrm{d}T}{\mathrm{d}r_{+}}.
\end{align}
We present the above two quantities in Appendix, as their forms are very complicated.

\section{$P-V$ criticality}
\label{PV}
According to Eq.(\ref{temperature}), we can obtain the the equation of state (EOS)
\begin{align}\label{EOS}
  &1024P^2\pi^2r_+^4+64k(d-1)(d-2)P\pi\,r_+^2+k^2d(d-1)^2(d-4)(1-\epsilon^2)\nonumber\\
  &-8(d-1)\pi\,r_+\bigg(32P\pi\,r_+^2+kd(d-1)(1-\epsilon^2)\bigg)T=0,
\end{align}
which reflects the double-valuedness of the pressure in the extended thermodynamic phase space.
On the other hand, we can derive the pressure $P(T,r_{+})$ as
\begin{align}
  P(T,r_{+})=\frac{(d-1)}{32\pi\,r_+^2}\bigg(-kd+2(k+2\pi\,r_+T)+\sqrt{4(k+2\pi\,r_+T)^2-kd\epsilon^2(4k-kd+8\pi\,r_+T)}\bigg),
\label{P(T,r)}
\end{align}
while another one is the negative pressure branch.
After taking a series expansion, it leads to
\begin{align}
  P=\frac{(d-1)T}{4r_+}-\frac{k(d-1)(d\epsilon^2+d-4)}{32\pi\,r_+^2}+\mathcal {O}(r_+^{-3}).
\end{align}
Comparing the above equation with the Van der Waals equation
\begin{align}
  P=\frac{T}{\nu-b}-\frac{a}{\nu^2}\simeq\frac{T}{\nu}+\frac{bT}{\nu^2}-\frac{a}{\nu^2}+\mathcal {O}(\nu^{-3}),
\end{align}
one can easily find the specific volume $\nu\propto\,r+$.
Therefore we will just use the horizon radius $r_+$ in EOS
instead of the specific volume $\nu$ and study the $P-r+$ behavior
in the following paper.

To consider the $P-V$ criticality, we can focus on
\begin{align}
  \frac{\partial P}{\partial r_+}=0,\quad\,\frac{\partial^2 P}{\partial r_+^2}=0,
\end{align}
to find the critical points.
As the direct differentiation of $P$ (Eq.\ref{P(T,r)}) is too complicated, we
prefer to employ the implicit differentiation on EOS (Eq.\ref{EOS}) and the above equations, and we can derive two simple conditions
\begin{align}
  &512P^2\pi\,r_{+}^3+16(d-1)\bigg(k(d-2)r_{+}-6\pi\,r_{+}^2T\bigg)P-kd(d-1)^2(1-\epsilon^2)T=0,\label{ceq1}\\
  &96P\pi\,r_+^2-12(d-1)\pi\,r_{+}T+(d-1)(d-2)k=0 \label{ceq2}.
\end{align}
They lead to the critical points:
\begin{align}
  P_c&=\frac{k(d-1)(12\pi\,X+2-d)}{96\pi\,r_+^2},\\
  T_c&=\frac{kX}{r_+},
\end{align}
by which, the EOS (Eq.\ref{EOS}) is simplified as a equation of $\epsilon$ and results in
the following condition
\begin{align}
  \epsilon_{c}=\pm2\sqrt{\frac{(2(d-5)X\pi+2d-7)}{3d(d-4)}},
\end{align}
where
\begin{align}
  X=\frac{3(d-3)\pm\sqrt{-3(d-1)(d-5)}}{12\pi}.
  \label{Xvalue}
\end{align}

It is interesting that there is no critical volume or horizon radius, but the critical parameter $\epsilon$.
To start the physical discussion, we should firstly calculate the physical critical points with real and positive
pressure and temperature, which lead to the constraint
\begin{align}
  1\leq\,d\leq5.
\end{align}
Namely, there is $P-V$ criticality only in four, five and six dimensions.
Besides, the physical critical points are simplified as:
\begin{equation}
\left\{ \begin{array}{lll}
P_c=\frac{k(2\sqrt{3}\gamma-1)}{48\pi\,r_+^2},\quad T_c=\frac{\sqrt{3}k\gamma}{6\pi\,r_+},\quad \epsilon_c=\pm\frac{2}{9}\sqrt{9+6\sqrt{3}\gamma},
\quad d=3; \\
P_c=\frac{k}{8\pi\,r_+^2},\qquad\quad T_c=\frac{k}{2\pi\,r_+},\quad \epsilon_c=\pm\frac{2}{3}\sqrt{2},
\qquad\qquad\quad  d=4; \\
P_c=\frac{k}{8\pi\,r_+^2},\qquad\quad T_c=\frac{k}{2\pi\,r_+},\quad \epsilon_c=\pm\frac{2}{5}\sqrt{5}, \qquad\qquad\quad d=5,
\end{array} \right.
\end{equation}
where $\gamma=\pm1$.
It is easy to find that positive pressure and temperature require the conditions $(k=1,\gamma=1)$.
In four dimensions ($d=3$), the critical behavior is studied in \cite{Ma:2017pap}.

Totally, we conclude that only four, five and six dimensional HL AdS black
holes with spherical horizon have physical critical points, as shown in Table.\ref{table}.

\begin{table}
\begin{center}
\begin{tabular}{|c|c|c|c|c|}
\hline
Dimensions &  $P_c$ & $T_c$ & $\epsilon_c$& critical relation $\frac{P_cr_+}{T_c}$ \\
\hline
Four &  $\frac{2\sqrt{3}-1}{48\pi\,r_+^2}$ & $\frac{\sqrt{3}}{6\pi\,r_+}$ & $\pm\frac{2}{9}\sqrt{9+6\sqrt{3}}$& $\frac{6-\sqrt{3}}{24}\approx0.178$ \\
\hline
Five &  $\frac{1}{8\pi\,r_+^2}$ & $\frac{1}{2\pi\,r_+}$ & $\pm\frac{2}{3}\sqrt{2}$ & $\frac{1}{4}$\\
\hline
Six &  $\frac{1}{8\pi\,r_+^2}$ & $\frac{1}{2\pi\,r_+}$ & $\pm\frac{2}{5}\sqrt{5}$ & $\frac{1}{4}$\\
\hline
\end{tabular}
\caption{The critical points of HL AdS spherical black holes.}\label{table}
\end{center}
\end{table}

In next sections, we will give the physical discussion about the critical  phenomena,
i.e. the infinitely critical points and continuous second order phase transitions.
Especially for four dimensional case, it is reported in \cite{Ma:2017pap},
which is called as ``peculiar critical phenomena". In Section.\ref{4+5D}, we will conclude that they are actually the famous $\lambda$ phase transition after studying the specific heat $C_P$ of HL black holes.
We also present the $\lambda$ phase transition in five dimensions.
In Section.\ref{6D}, it is found that six dimensional HL AdS black holes only contain ``normal" black hole phase, thus no similar phase transition.
In Section.\ref{critical phenomena-1} and Section.\ref{critical phenomena-2},
we discuss the critical phenomena and calculate the critical exponents in different planes, by identifying parameter $\epsilon$ as the ordering field instead of pressure and temperature, respectively.

\section{$\lambda$ phase transition in four and five dimensions}\label{4+5D}
\begin{figure}[h!]
\begin{center}
\includegraphics[width=0.4\textwidth]{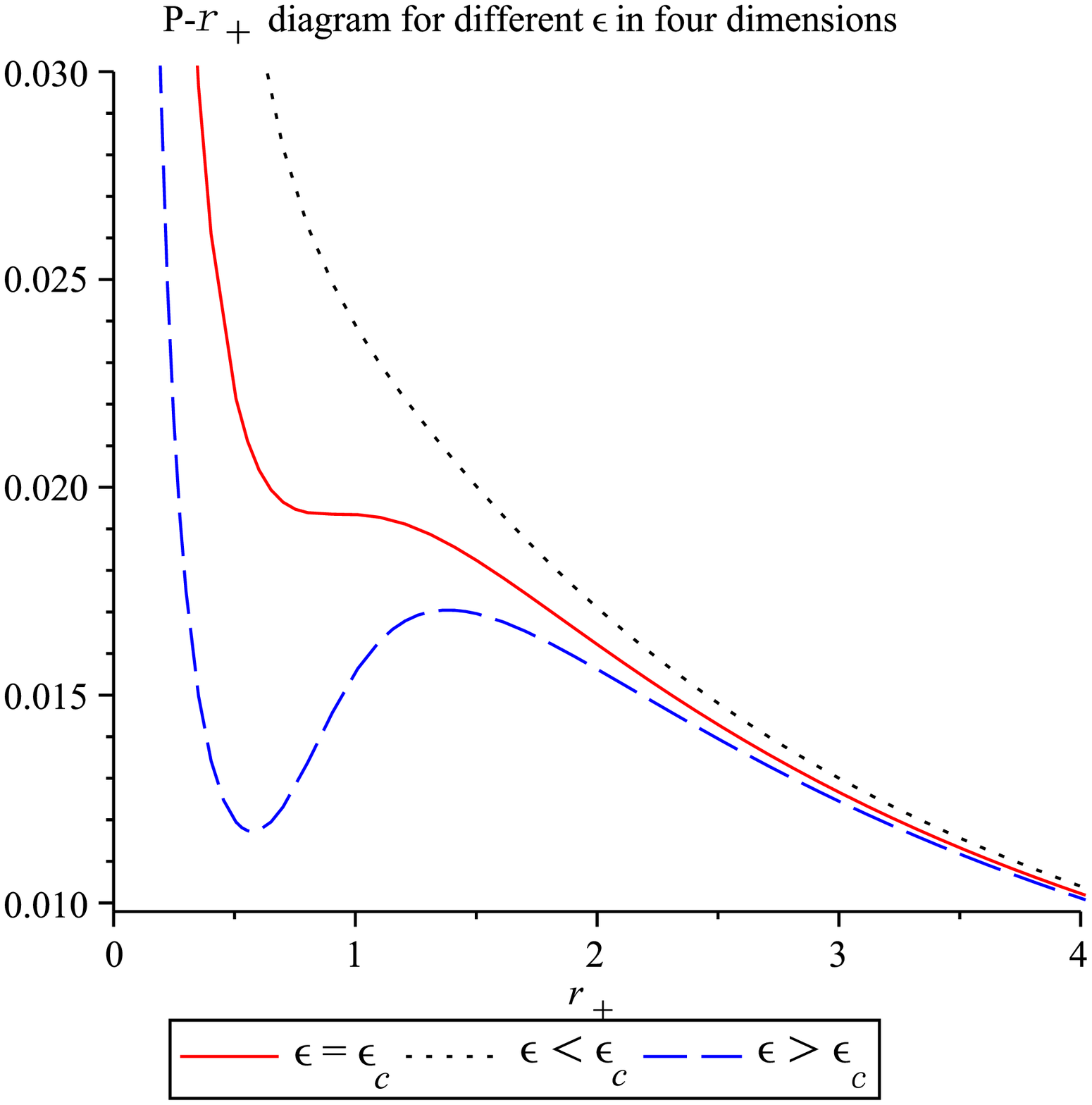}
\includegraphics[width=0.4\textwidth]{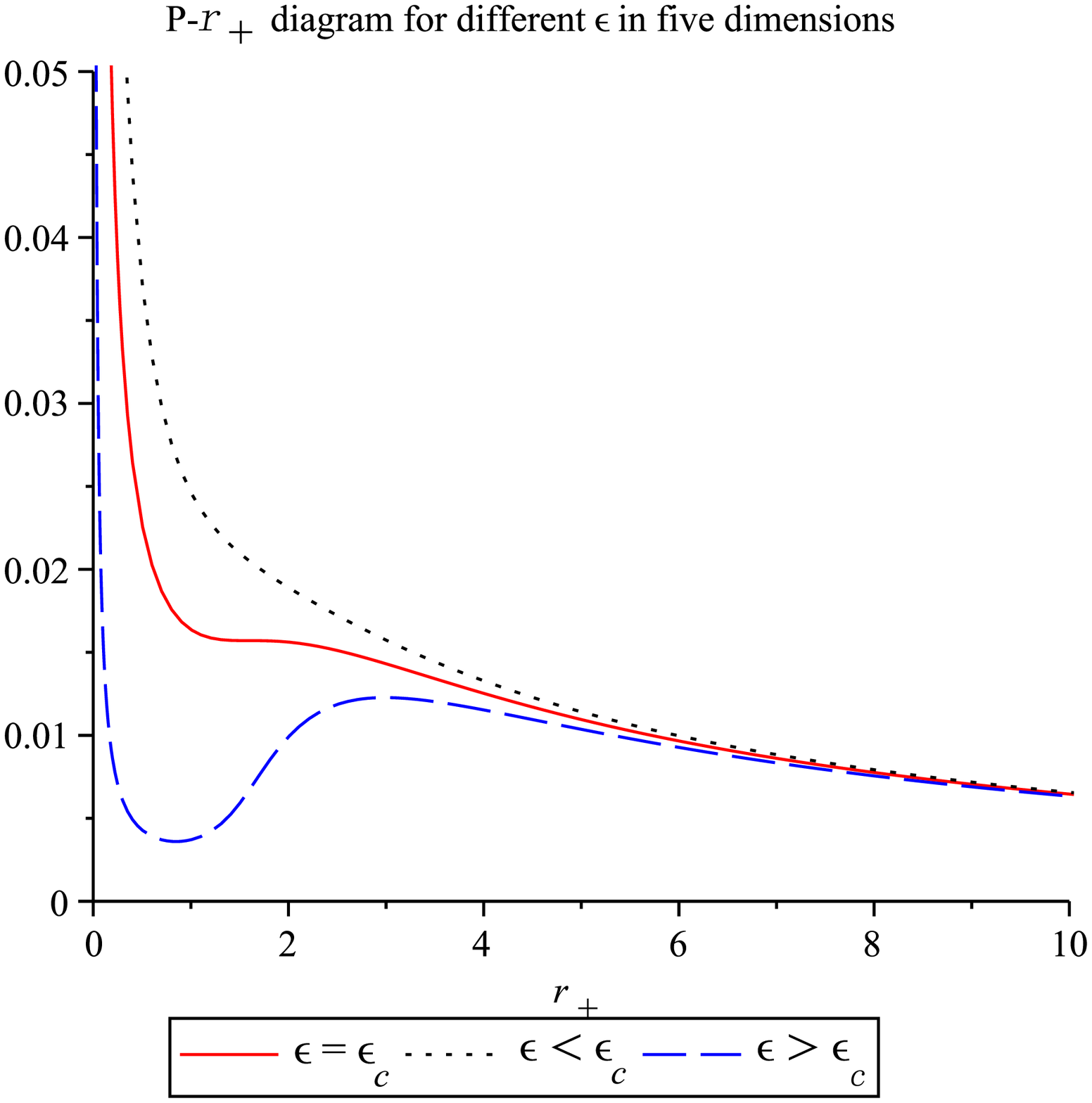}
\includegraphics[width=0.4\textwidth]{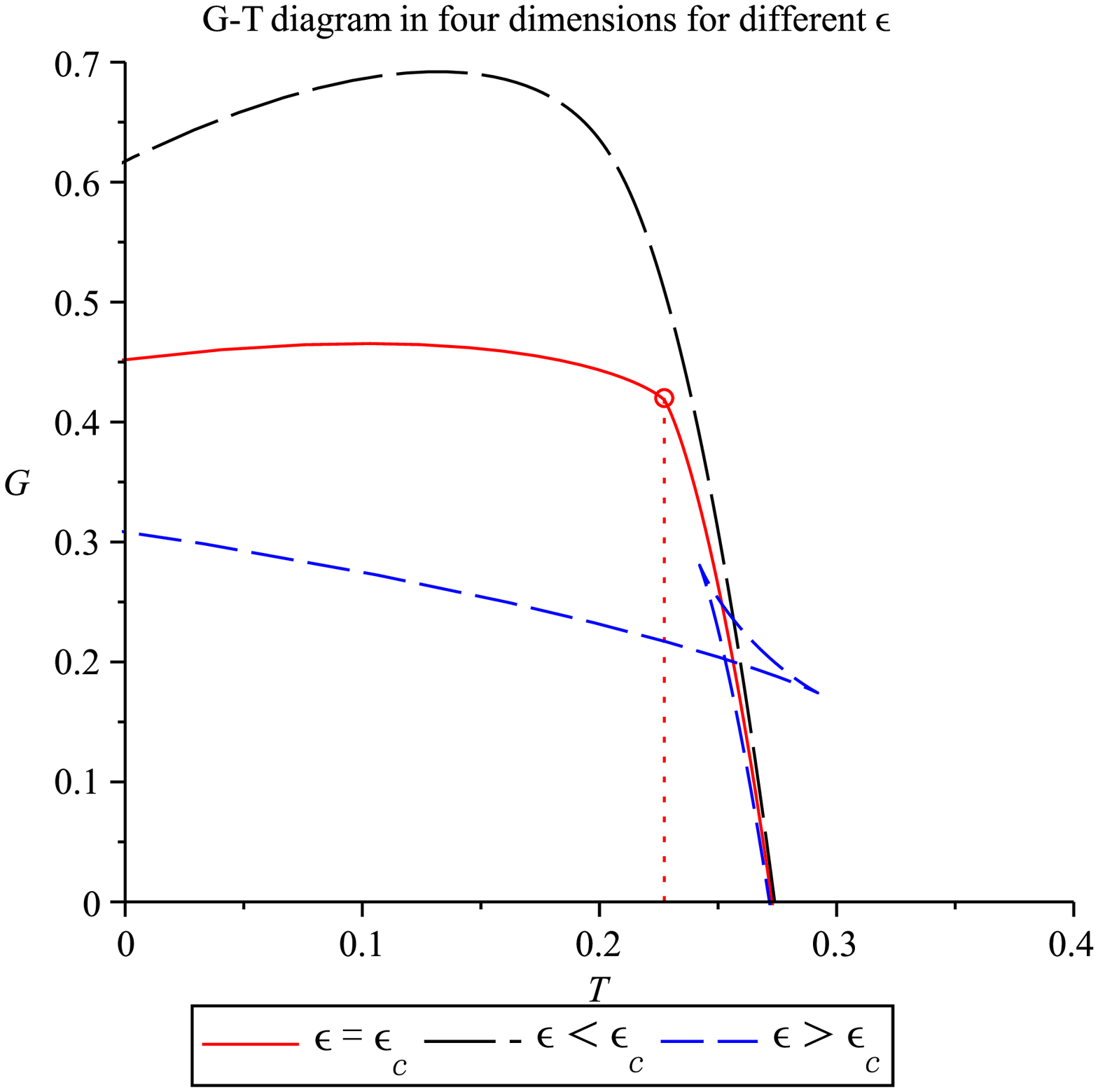}
\includegraphics[width=0.4\textwidth]{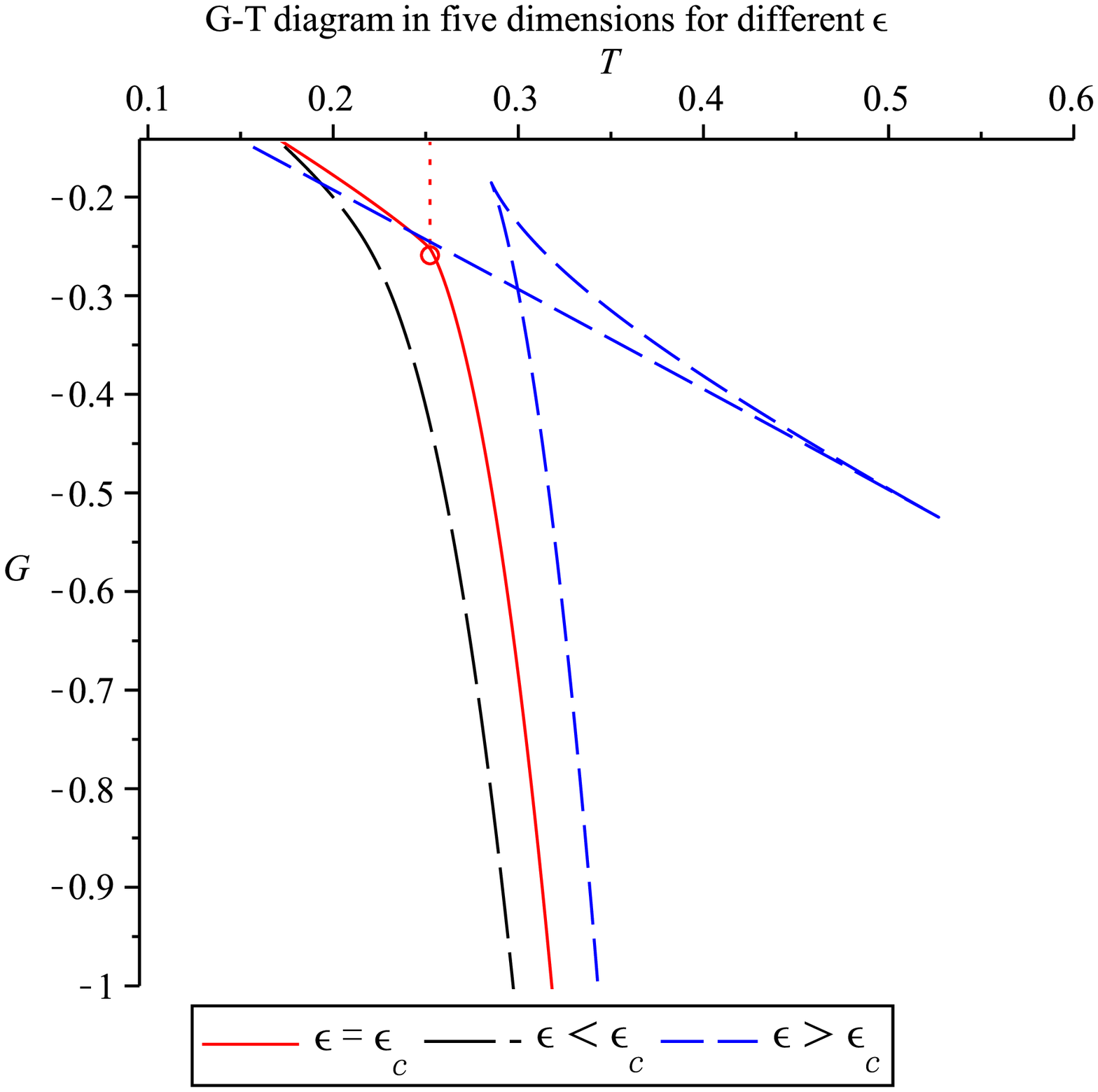}
\caption{Curves of $P-r_+$ at the same temperature $T=\frac{1}{10}$ and $G-T$ at the same pressure $P=\frac{1}{10}$ in four and five dimensions for different $\epsilon$.
When $\epsilon>\epsilon_c$, there are the $P-r_+$ oscillatory behavior and the classical ``swallow tail" characterizing the small/large black holes phase transition.
When $\epsilon=\epsilon_c$, the second order phase transition emerges and the dotted lines highlight the points where the second derivative of the Gibbs free energy $G$ diverges.
}
\label{4D5D-1}
\end{center}
\end{figure}

In this section, we discuss phase transitions in four and five dimensions.
Back to the critical points in Table.\ref{table},
it is interesting that there is no critical volume or horizon radius, but the critical parameter $\epsilon$.
Moreover, the $P-r_+$ oscillatory behavior and the classical ``swallow tail" characterizing the first order phase transition are controlled by the parameter $\epsilon$, instead of the temperature $T$, which is clearly shown in Fig.\ref{4D5D-1}. The $P-r_+$ diagrams are plotted at the same temperature $T$ and $G-T$ diagrams are plotted at the same pressure $P$. When $\epsilon>\epsilon_c$, there exist the small/large black holes phase transition in both four and five dimensions, which is exactly the same as the liquid/gas  phase transition of van der Waals fluid.
Especially for $\epsilon=\epsilon_c$, the $P-r_+$ curves become a critical
isotherm having an inflection point, and the second order phase transition emerges.

\begin{figure}[h!]
\begin{center}
\includegraphics[width=0.4\textwidth]{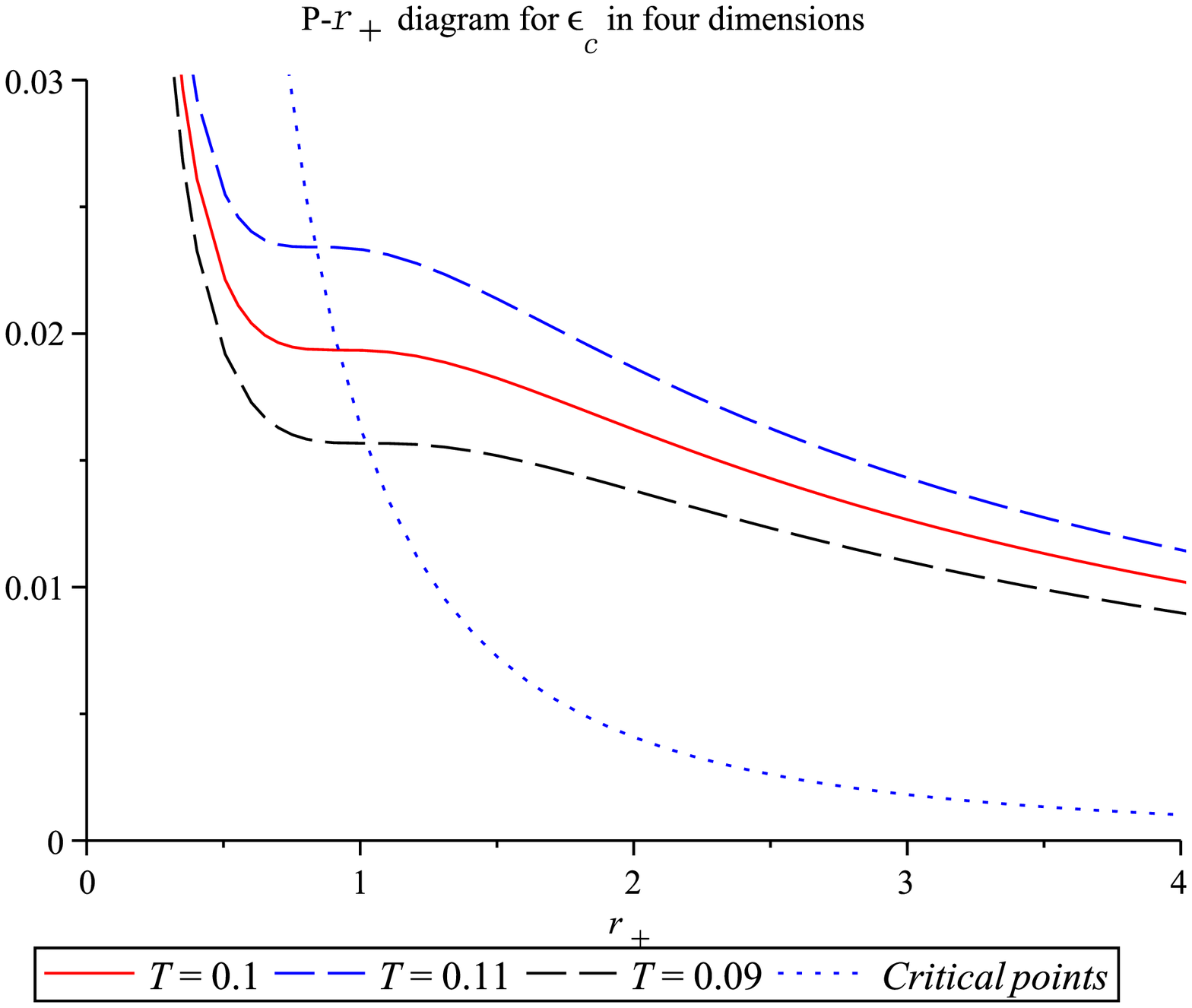}
\includegraphics[width=0.4\textwidth]{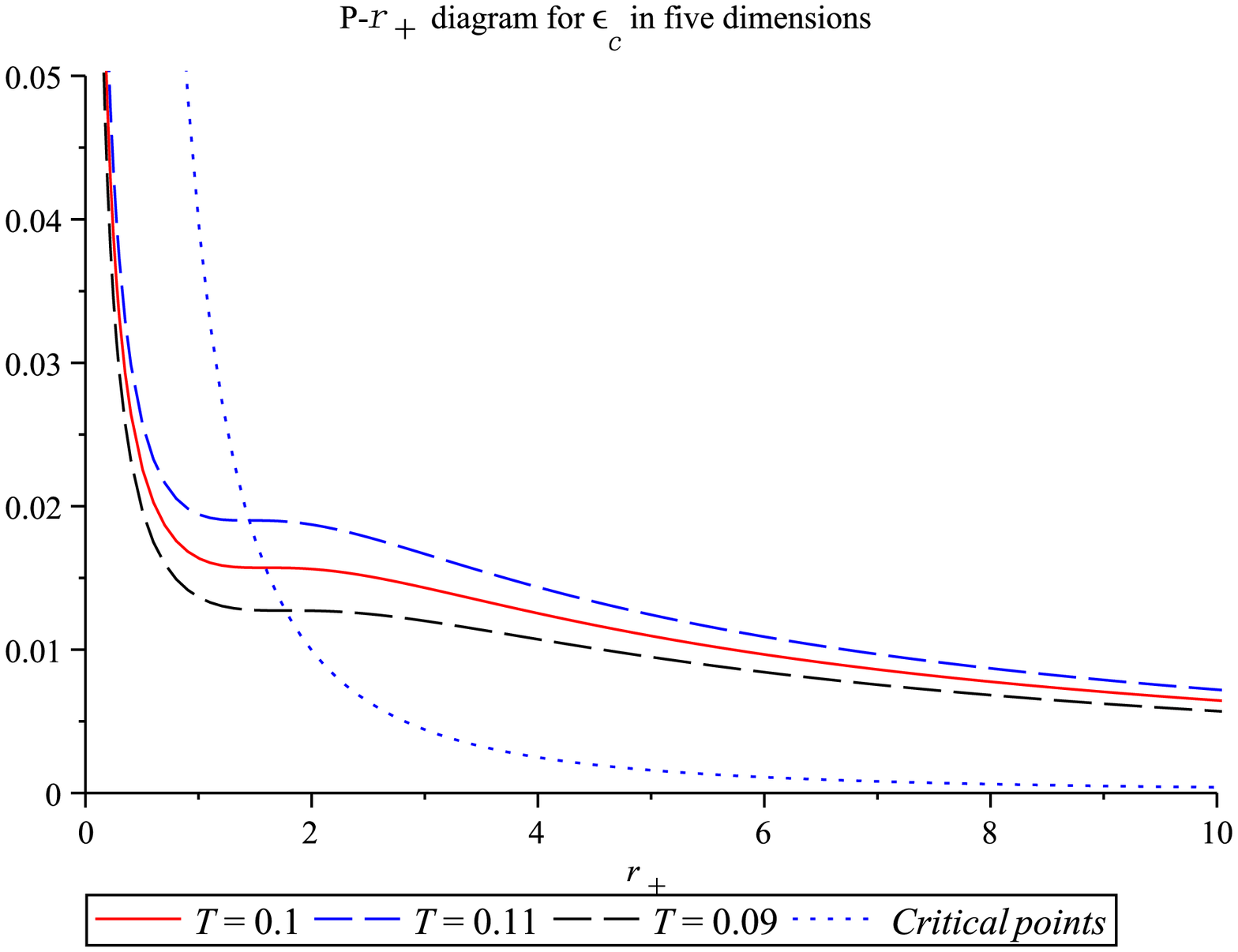}
\includegraphics[width=0.4\textwidth]{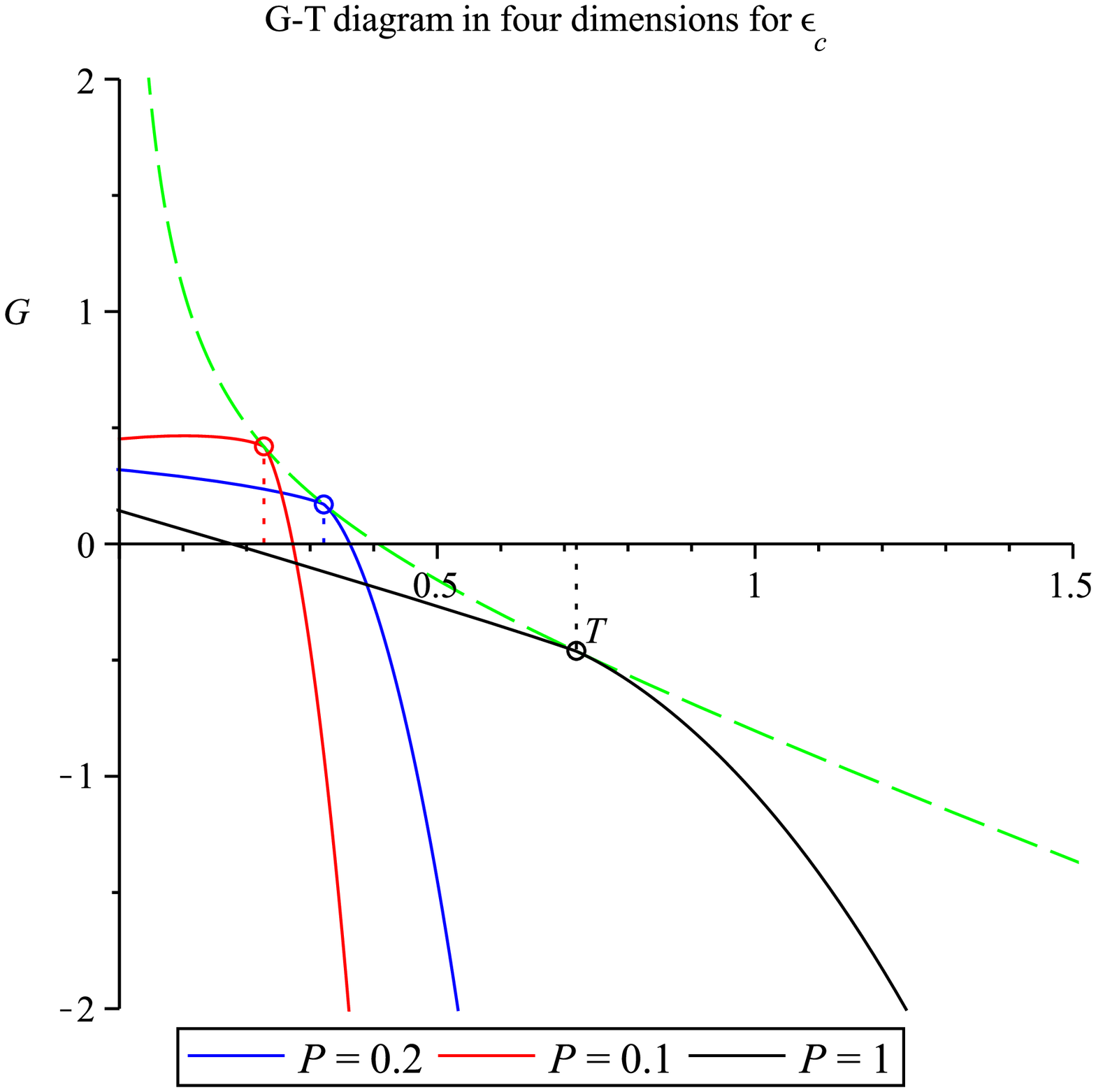}
\includegraphics[width=0.4\textwidth]{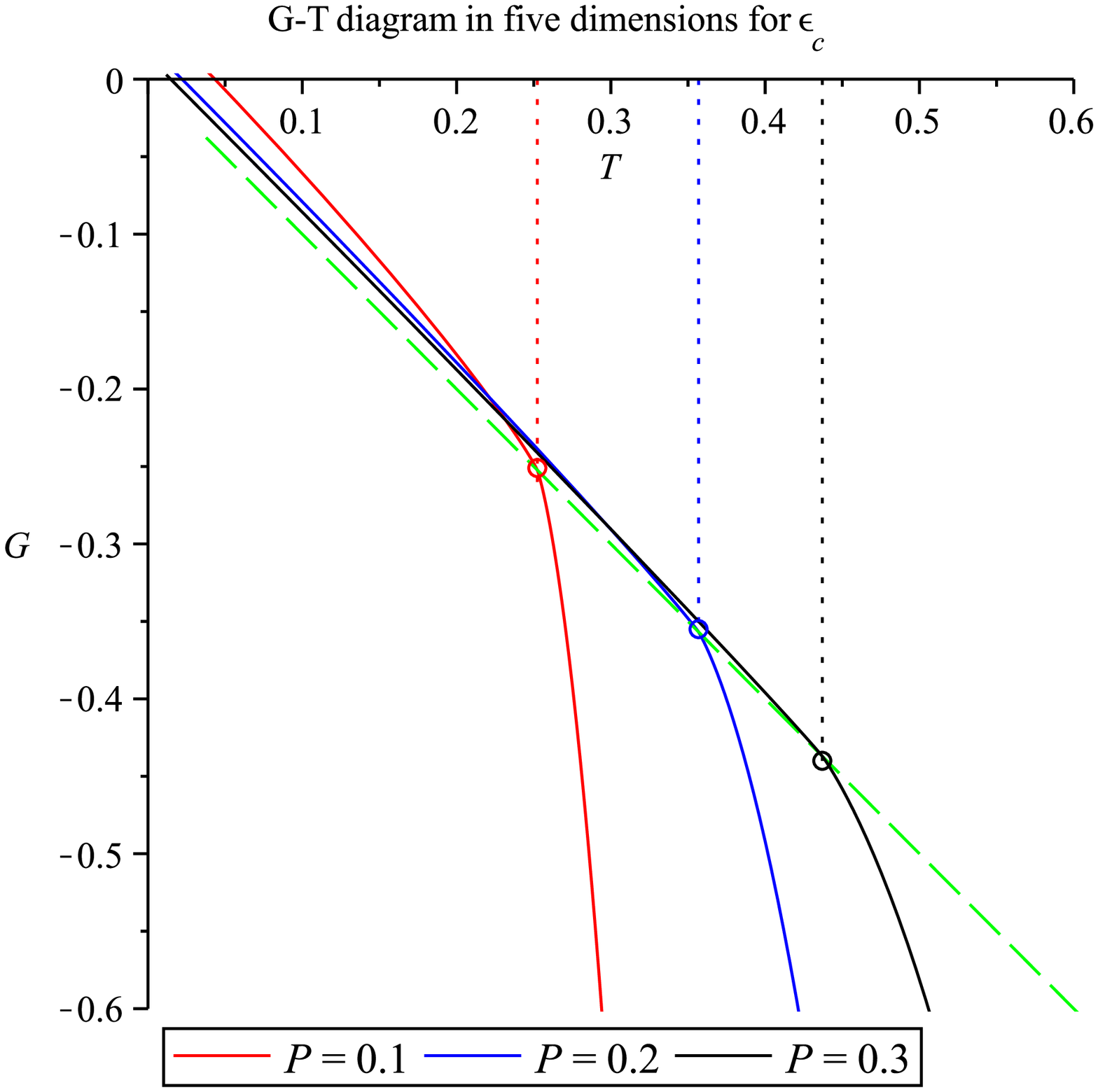}
\caption{Curves of $P-r_+$ at the different temperature $T$ and $G-T$ at the different pressure $P$ in four and five dimensions for $\epsilon=\epsilon_c$.
Every isotherm in $P-r_{+}$ diagrams is a critical
isotherm, and the dotted lines describe
the position of all critical points.
A line of second order phase transitions are shown in $G-T$ diagrams, and
the dashed lines describe the position of all phase transition points.
}
\label{4D5D-2}
\end{center}
\end{figure}

On the other hand, black holes with spherical horizon and arbitrary mass $M$ always have event horizon $r_+>0$, which is arbitrary as well. As a result, four and five dimensional HL AdS black holes with $\epsilon=\epsilon_c$ exhibit infinitely many critical points with arbitrary temperature $T_c$ and horizon radius $r_+$ as
\begin{equation}\label{Plambda}
\left\{ \begin{array}{lll}
P_c=\frac{(2\sqrt{3}-1)\pi}{4}T_c^2,\quad T_c=\frac{\sqrt{3}}{6\pi\,r_+},
\quad d=3; \\
P_c=\frac{\pi}{2}T_c^2,\qquad\qquad T_c=\frac{1}{2\pi\,r_+},
\quad  d=4.
\end{array} \right.
\end{equation}
Namely, every isotherm in $P-r_{+}$ diagrams is a critical
isotherm, which has an inflection point at
  \begin{equation}\label{rcri}
 r_+=\left\{ \begin{array}{lll}
\frac{\sqrt{3}}{6\pi\,T},
\quad d=3; \\
\frac{1}{2\pi\,T},
\quad  d=4.
\end{array} \right.
\end{equation}
One can find it easily in Fig.\ref{4D5D-2}, and the dotted lines in $P-r_{+}$ diagrams describe
the position of all critical points (Eq.\ref{rcri}) at arbitrary temperature $T$.
Besides, in $G-T$ diagrams, there is no first order phase transition but
rather a line of second order phase transitions, for which the phase transition points are highlighted
by the dotted lines in Fig.\ref{4D5D-2}.
The dashed lines in $G-T$ diagrams describe the position of all phase transition points
 \begin{equation}\label{Tlambda}
T=   \left\{ \begin{array}{lll}
\sqrt{\frac{4P}{(2\sqrt{3}-1)\pi}},
\quad d=3; \\
\sqrt{\frac{2P}{\pi}},
\qquad\quad  d=4
\end{array} \right.
\end{equation}
at arbitrary pressure $P$, where the second derivative of the Gibbs free energy $G$ diverges.

\begin{figure}[h!]
\begin{center}
\includegraphics[width=0.31\textwidth]{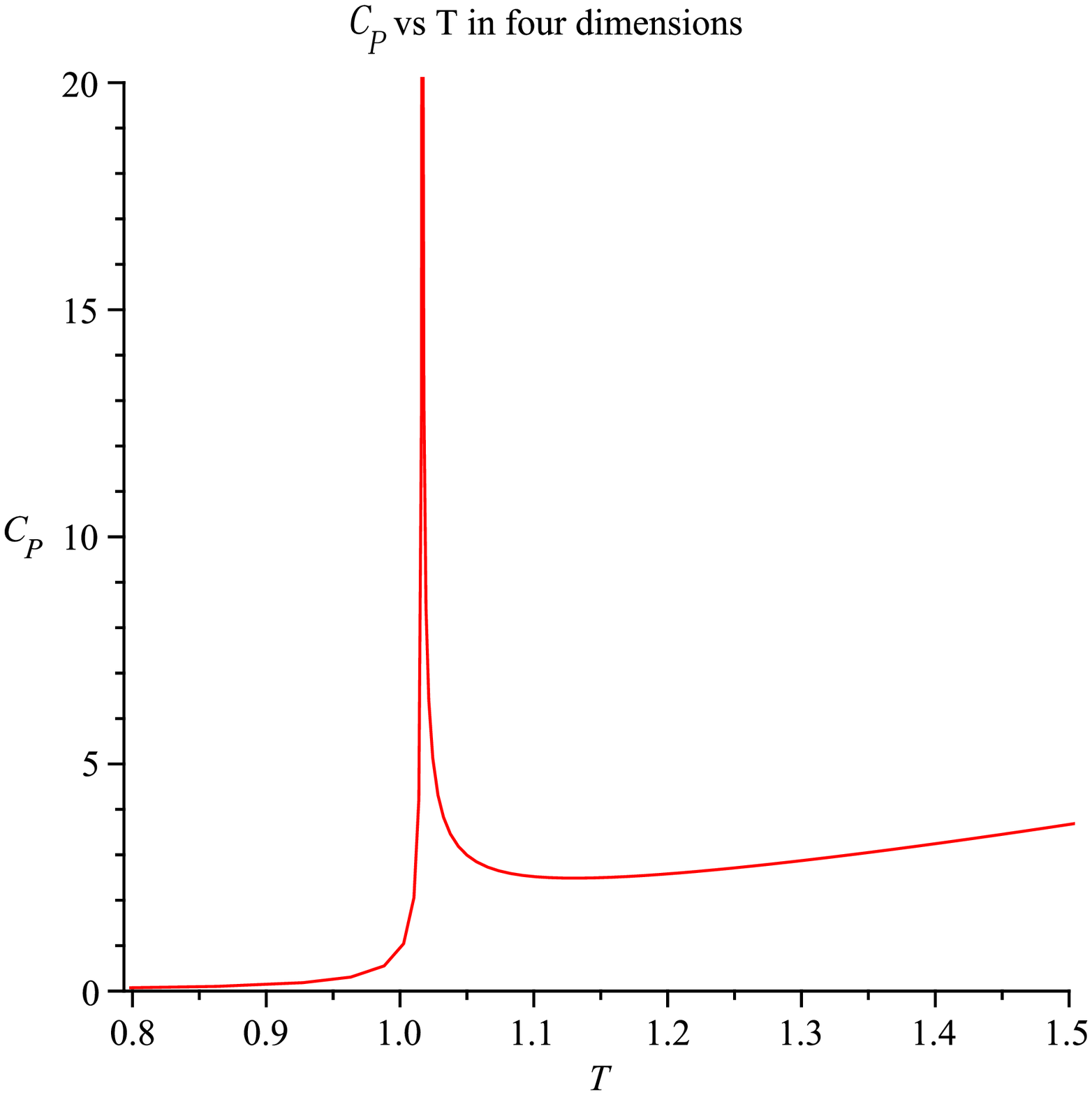}
\includegraphics[width=0.31\textwidth]{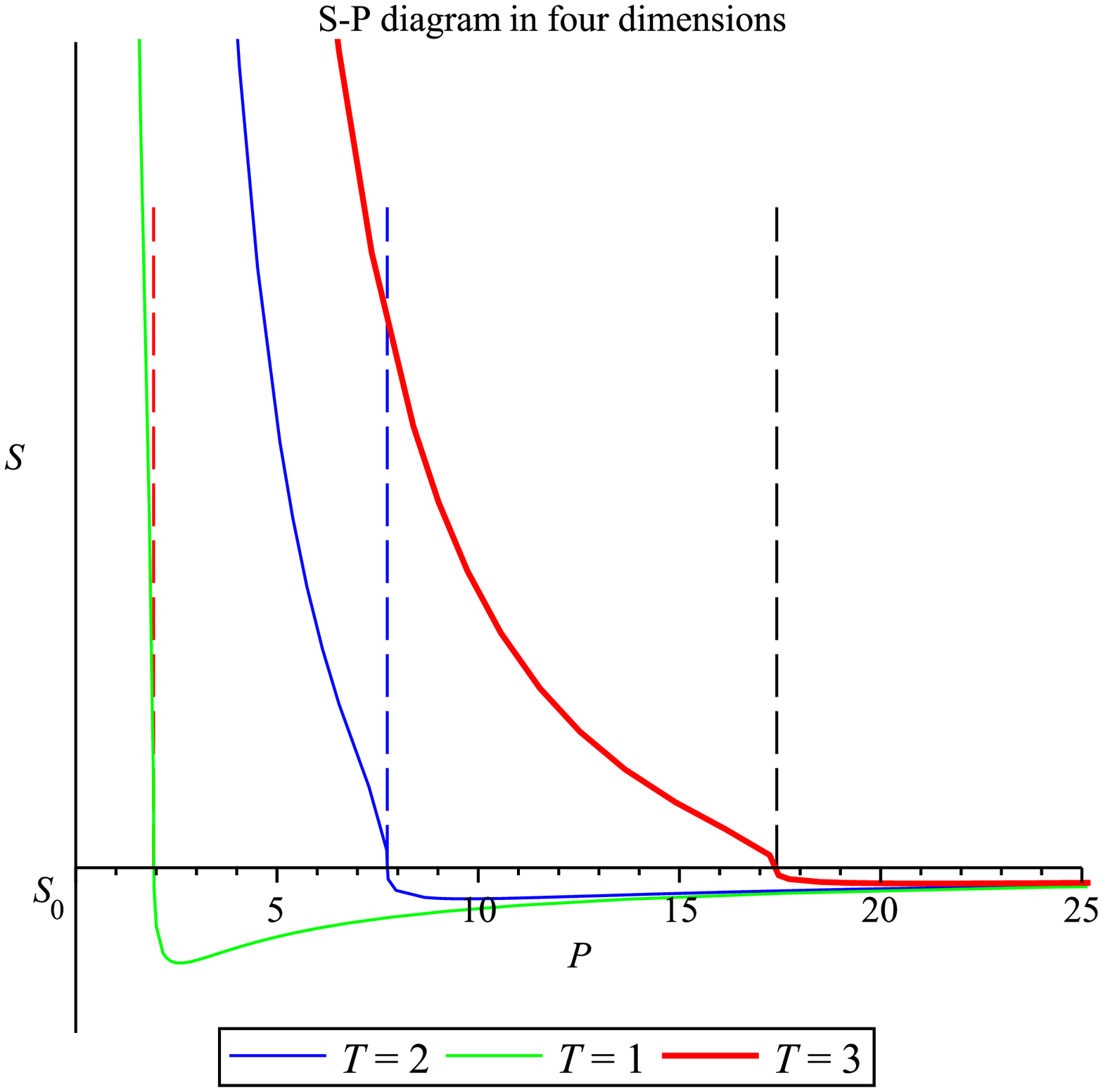}
\includegraphics[width=0.31\textwidth]{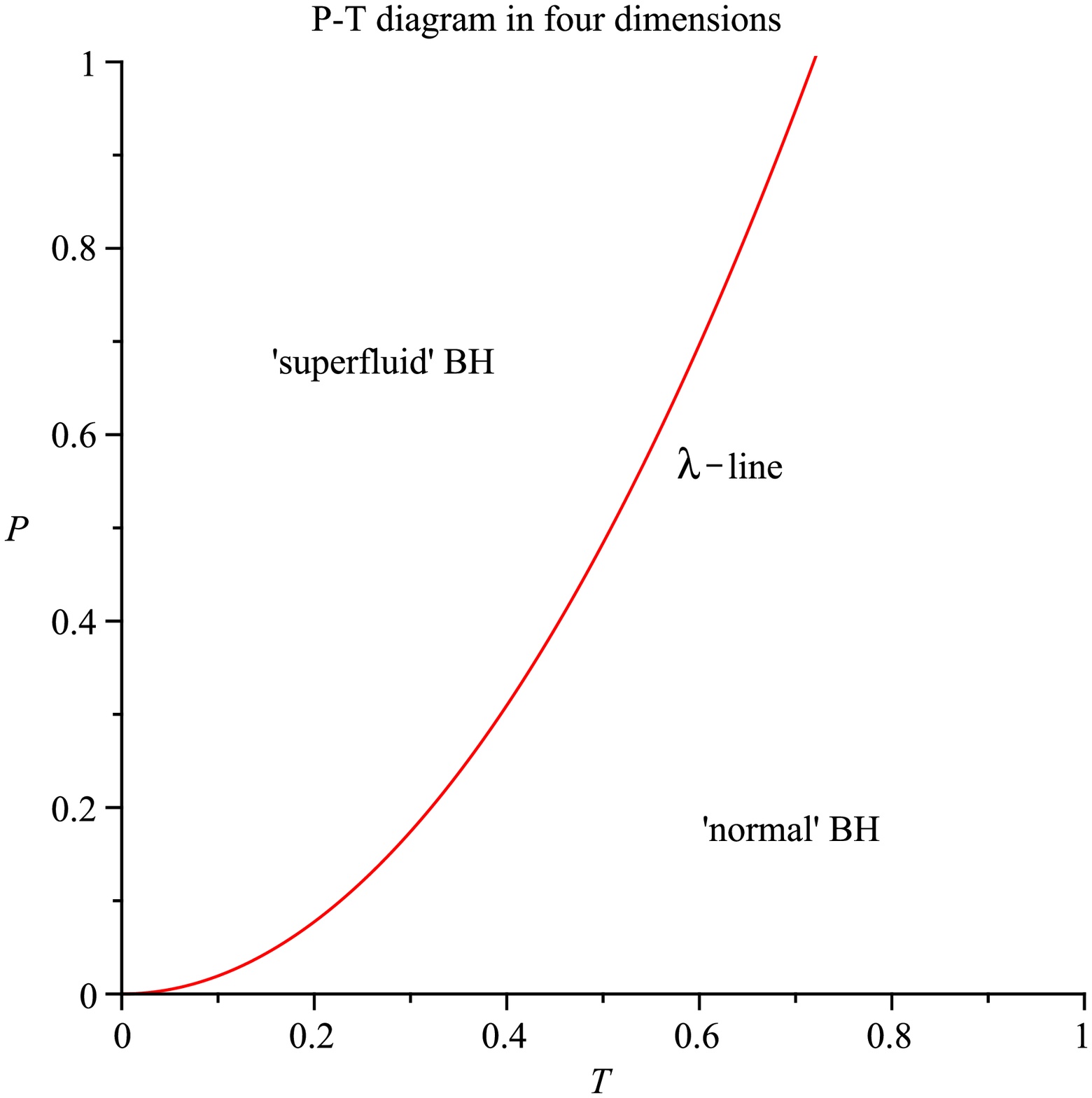}
\includegraphics[width=0.31\textwidth]{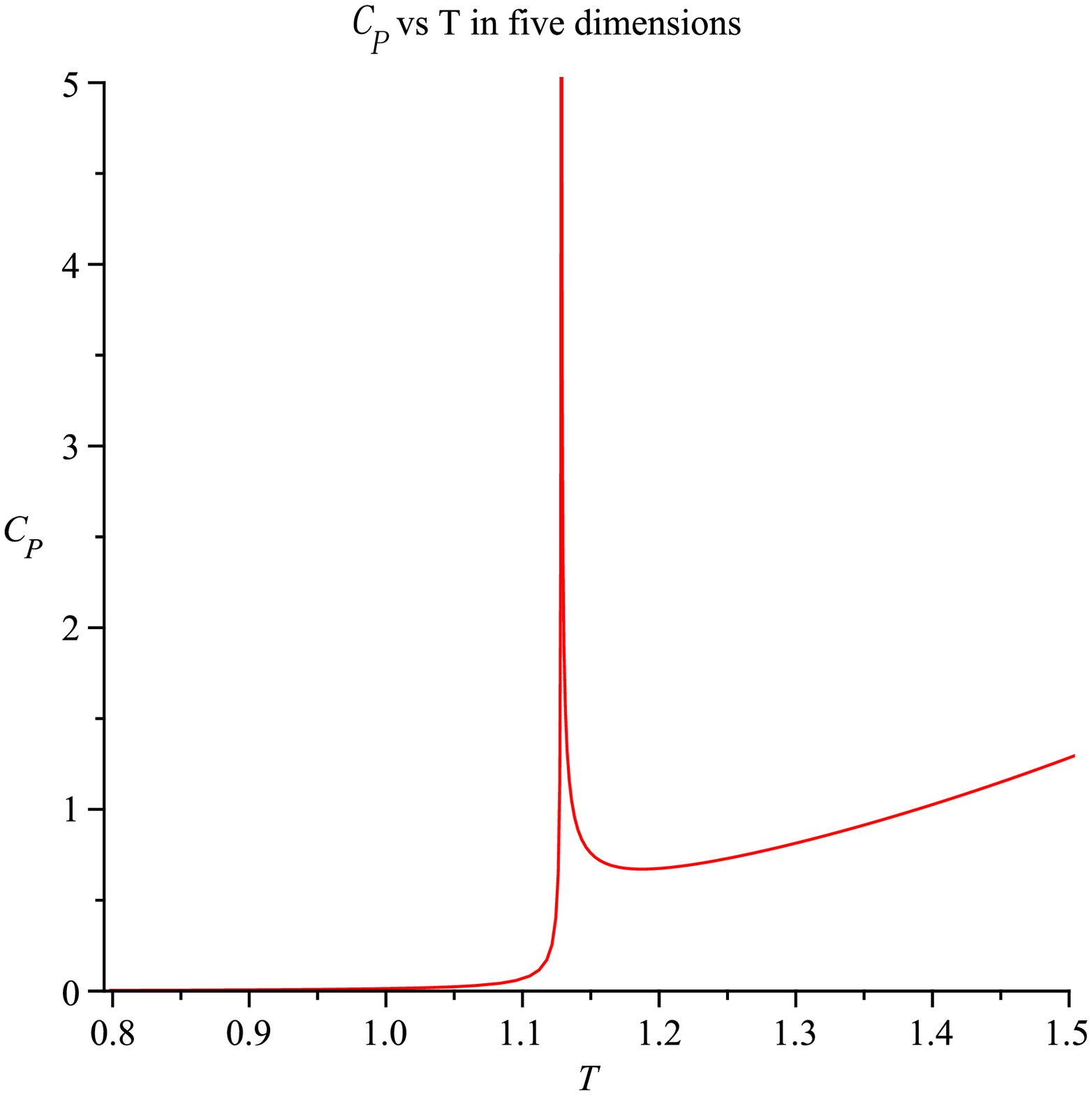}
\includegraphics[width=0.31\textwidth]{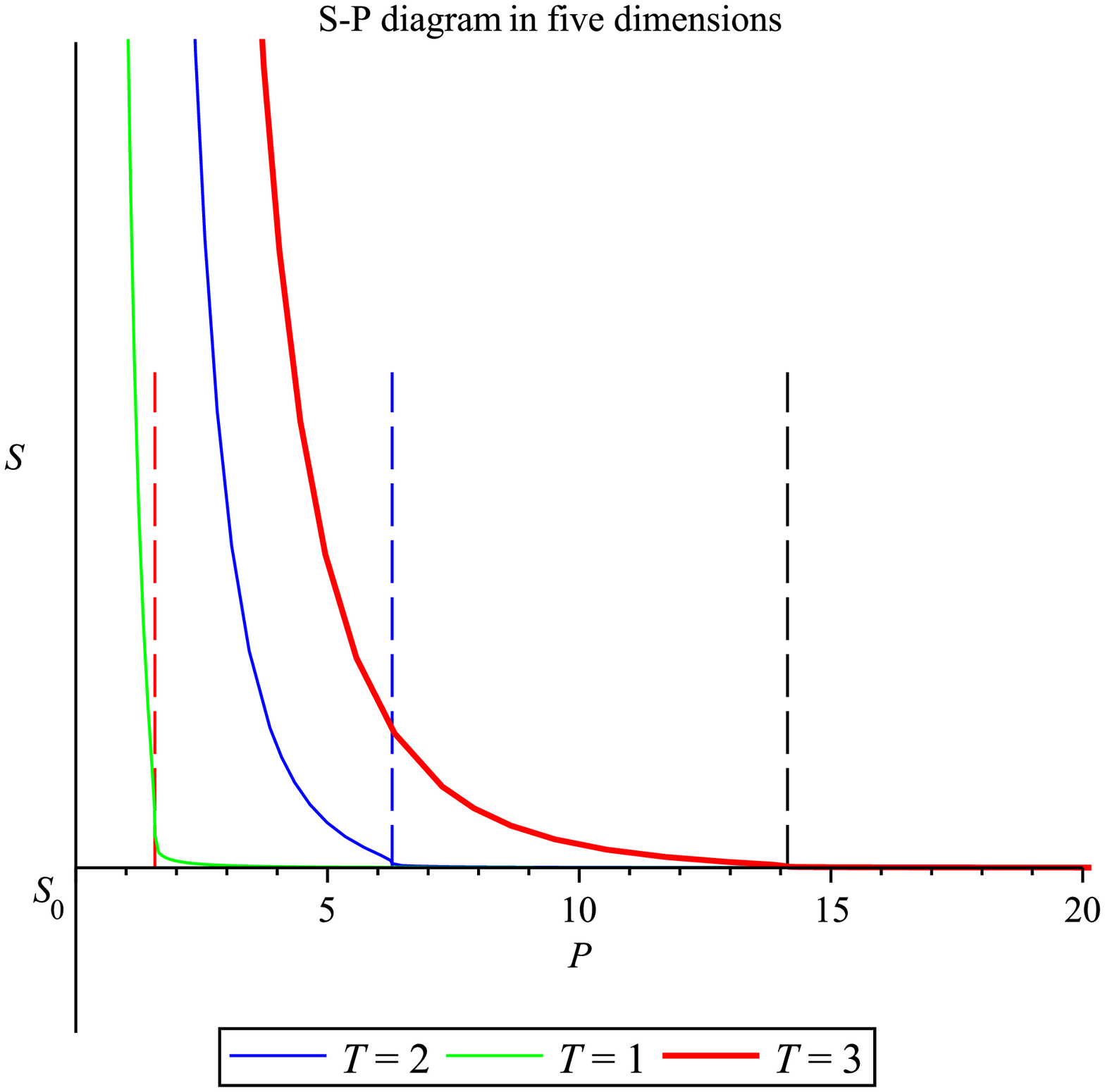}
\includegraphics[width=0.31\textwidth]{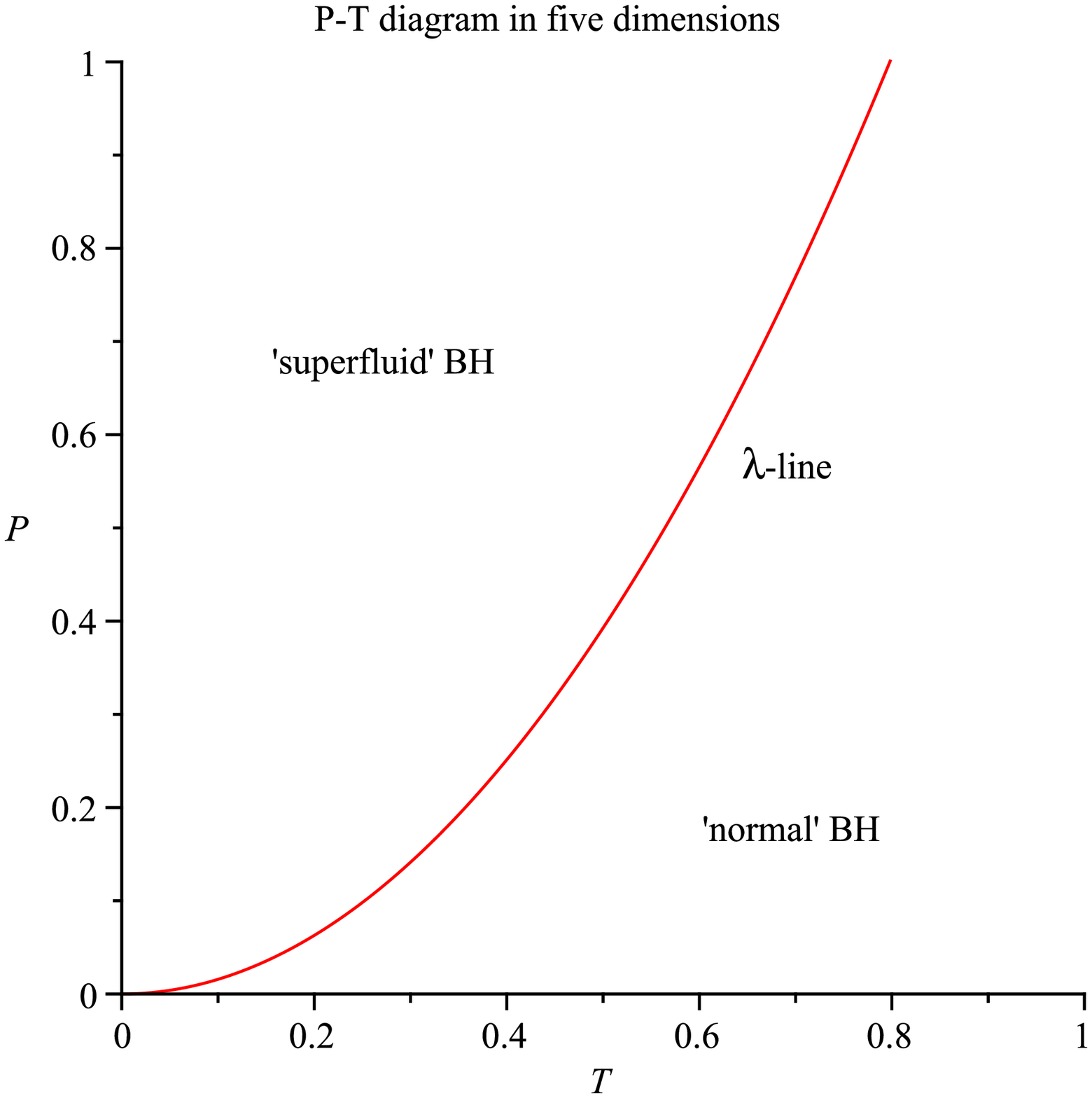}
\caption{
 Curves of $C_P-T$ and $S-P$, and $P-T$ parameter space of black holes with $\epsilon=\epsilon_c$ in four and five dimensions. The $\lambda$ line in $C_P-T$ and $P-T$ plots indicate the second order phase transitions between ``superfluid" black hole phase and  ``normal" black hole phase, which are distinguished in the $S-P$ plots by the dashed lines.
}
\label{4D5D-3}
\end{center}
\end{figure}

To study the continuous second order phase transitions, we focus on the
specific heat $C_P$ of black holes with $\epsilon=\epsilon_c$.
From the $C_P-T$ diagrams in Fig.\ref{4D5D-3}, one can obtain that
the specific heat $C_P$ always diverges at the
critical temperature (Eq.\ref{Tlambda}).
This is the classical $\lambda$ line, i.e. the line
of second order phase transitions, in $C_P-T$ diagrams, which was famous in the discussion of superfluidity of liquid ${}^4$He.
Similarly, in the $P-T$ parameter space (right two plots of Fig.\ref{4D5D-3}), a line of critical points, i.e. the $\lambda$ line, separates the
two phases of system, which are the ``superfluid" black hole phase and  ``normal" black hole phase.
To determine the two phases, we consider the $S-P$ diagrams, i.e. the middle two plots in Fig.\ref{4D5D-3}.
The dashed lines highlight the critical pressure (Eq.\ref{Plambda}). One can observe that the entropy of black holes with pressure smaller than the critical pressure are bigger, which correspond to the ``normal" black hole phase.
Another one is the ``superfluid" black hole phase having smaller entropy and corresponding to black holes with pressure larger than the critical pressure. Especially, the entropy of ``superfluid" black hole phase is almost vanishing in five dimensions, while it seems to be negative in four dimensions.
Note that the positivity of entropy depends on $S_0$,
which is not clear and could be fixed by counting microscopic degrees of freedom in quantum theory
of gravity as argued in \cite{Cai:2009pe}.

Finally, we conclude that this continuous second order phase transition between small/large black holes corresponds to the phase transition between ``superfluid" black hole and ``normal" black hole, which is firstly reported in Lovelock gravity with conformally coupled scalar field \cite{Hennigar:2016xwd}.

\section{``Normal" black hole phase in six dimensions}\label{6D}
\begin{figure}[h!]
\begin{center}
\includegraphics[width=0.4\textwidth]{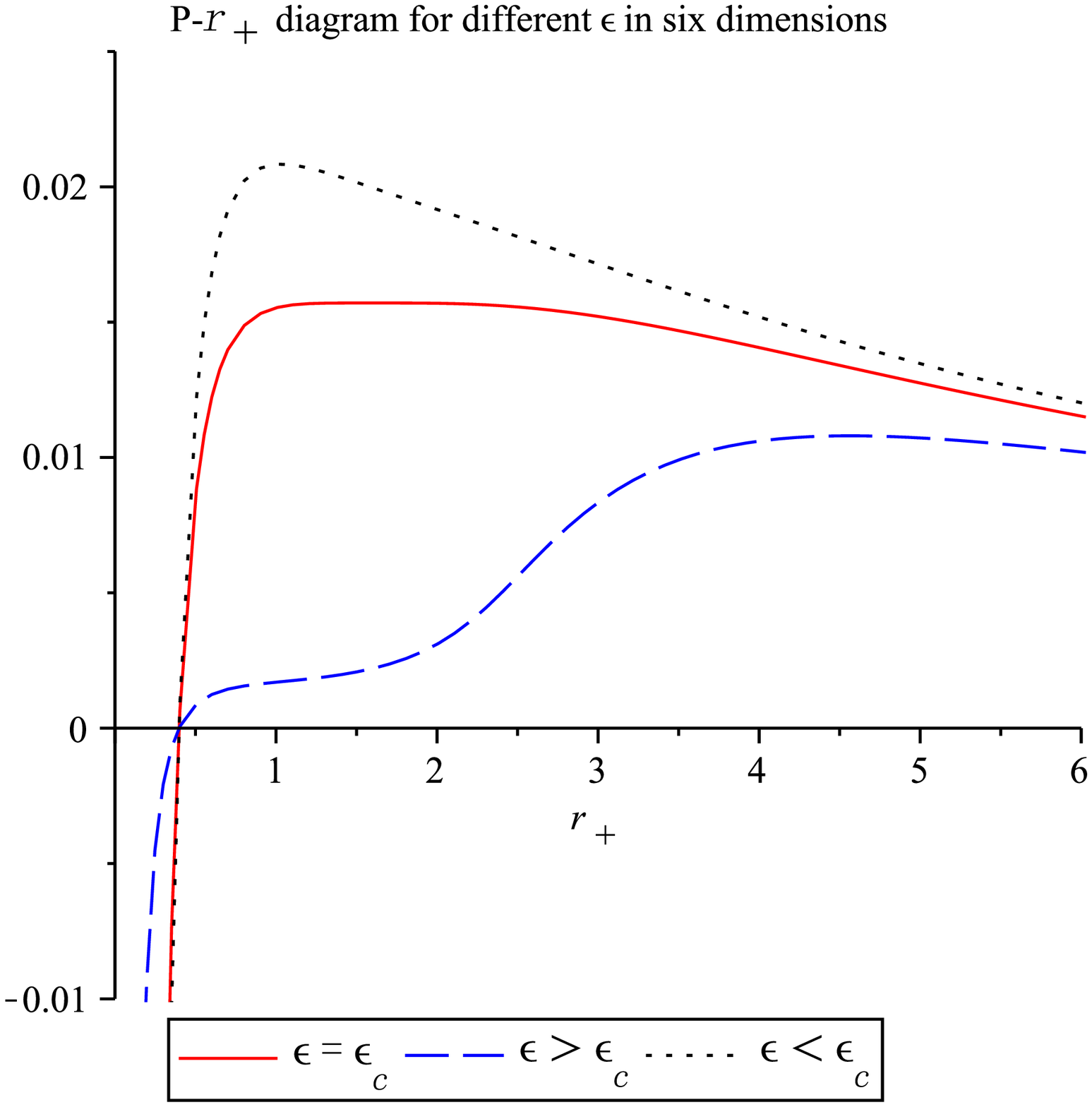}
\includegraphics[width=0.4\textwidth]{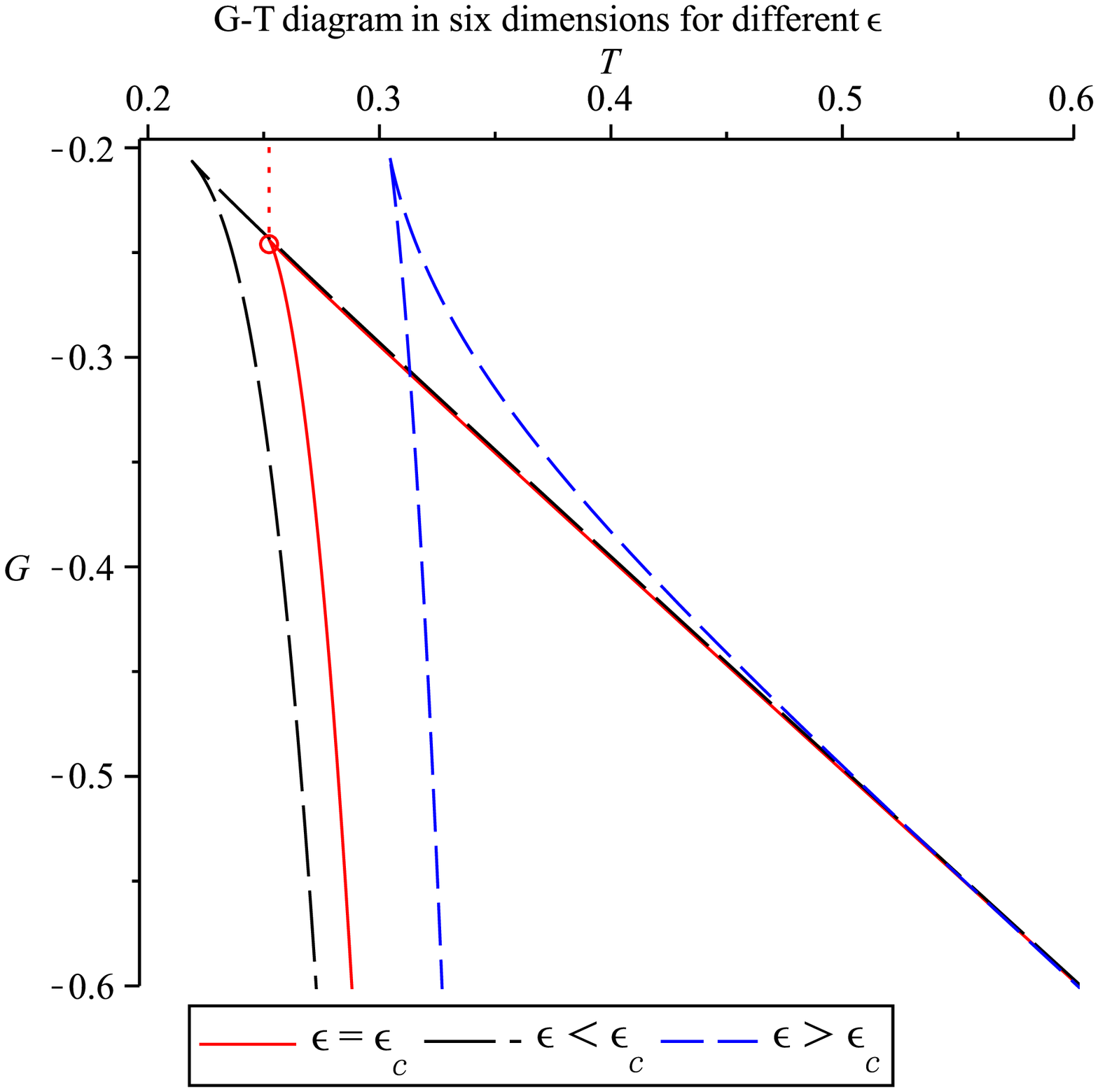}
\includegraphics[width=0.4\textwidth]{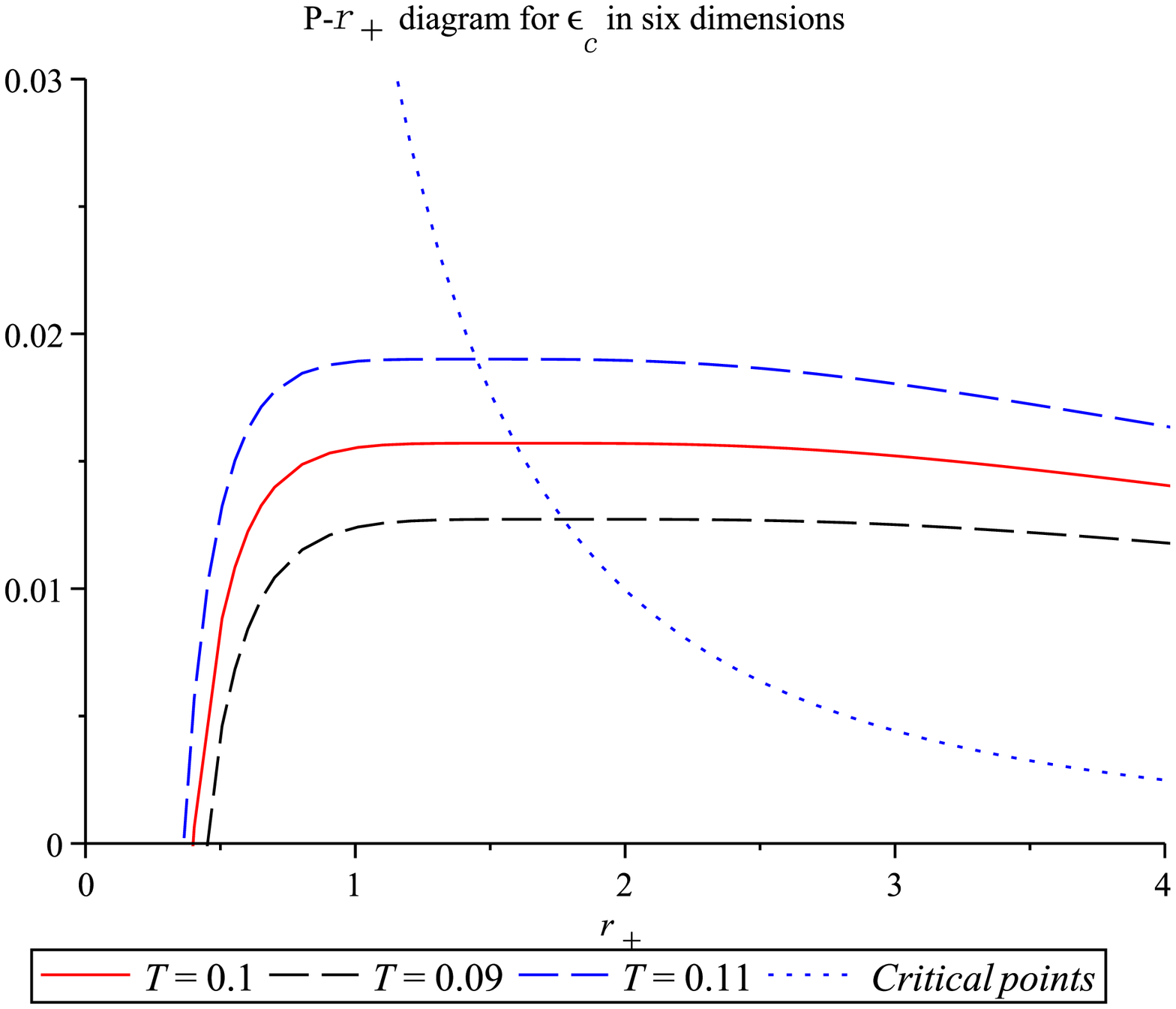}
\includegraphics[width=0.4\textwidth]{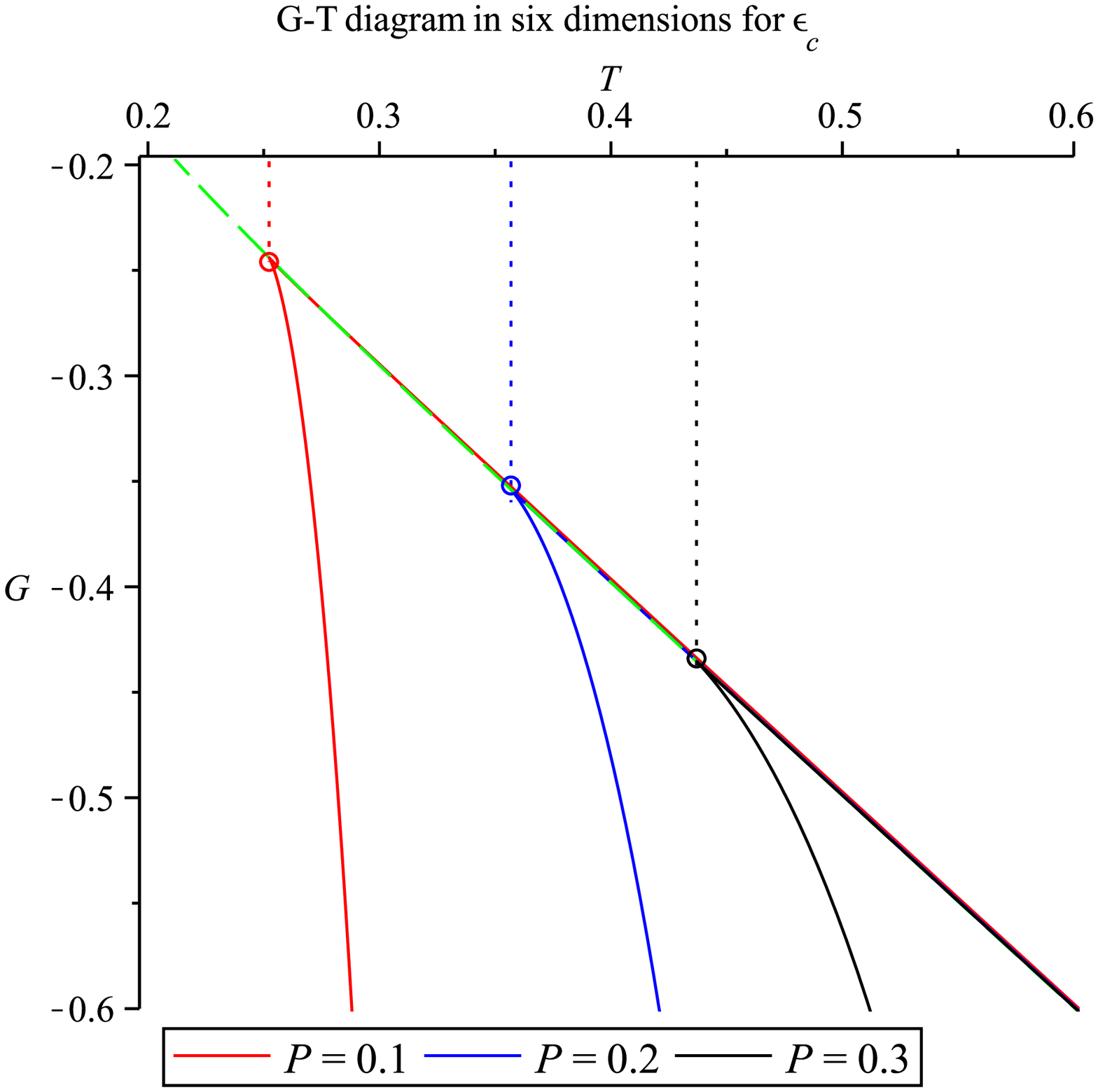}
\caption{Curves of $P-r_+$ and $G-T$ in six dimensions for different $\epsilon$ and $\epsilon=\epsilon_c$.
There is no $P-r_+$ oscillatory behavior and phase transition.
}
\label{6D-3}
\end{center}
\end{figure}

In six dimensions, there are critical points in $P-r_+$ plane as presented in Table.\ref{table}.
Actually, they describe infinitely many critical points with arbitrary temperature $T_c$ and horizon radius $r_+$
for HL black holes with $\epsilon=\epsilon_c=\pm\frac{2}{5}\sqrt{5}$, i.e.
\begin{align}
  P_c=\frac{\pi}{2}T_c^2,\quad\,T_c=\frac{1}{2\pi\,r_+},
\end{align}
which is exactly the same with the one in five dimensions.
However, from Fig.\ref{6D-3}, it is strange that there is no $P-r_+$ oscillatory behavior and the classical ``swallow tail" for different $\epsilon$; especially for $\epsilon=\epsilon_c$, one can never find the second order phase transitions. This is different from the cases in four and five dimensions, and six dimensional black holes seem to have no phase transition.

\begin{figure}[h!]
\begin{center}
\includegraphics[width=0.31\textwidth]{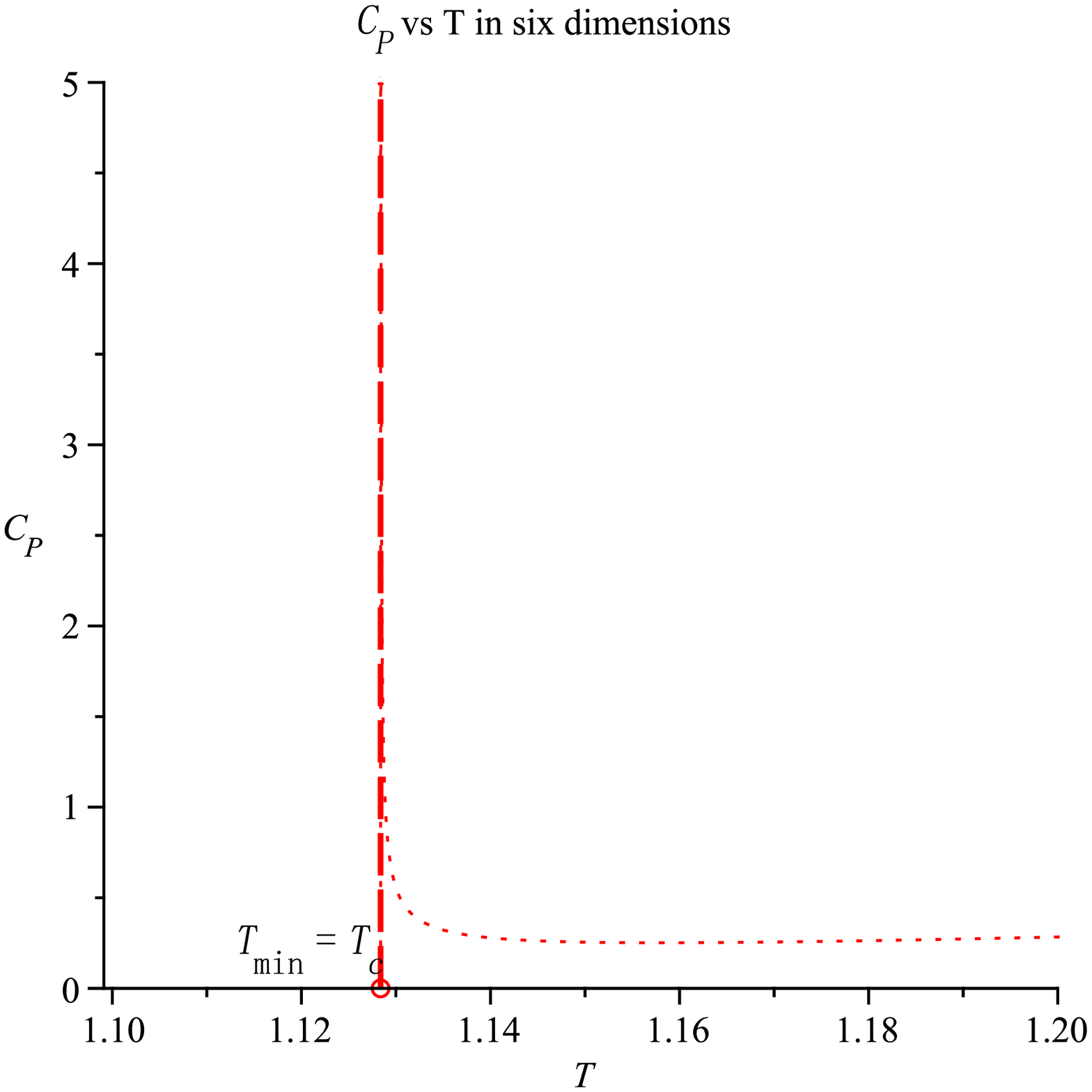}
\includegraphics[width=0.31\textwidth]{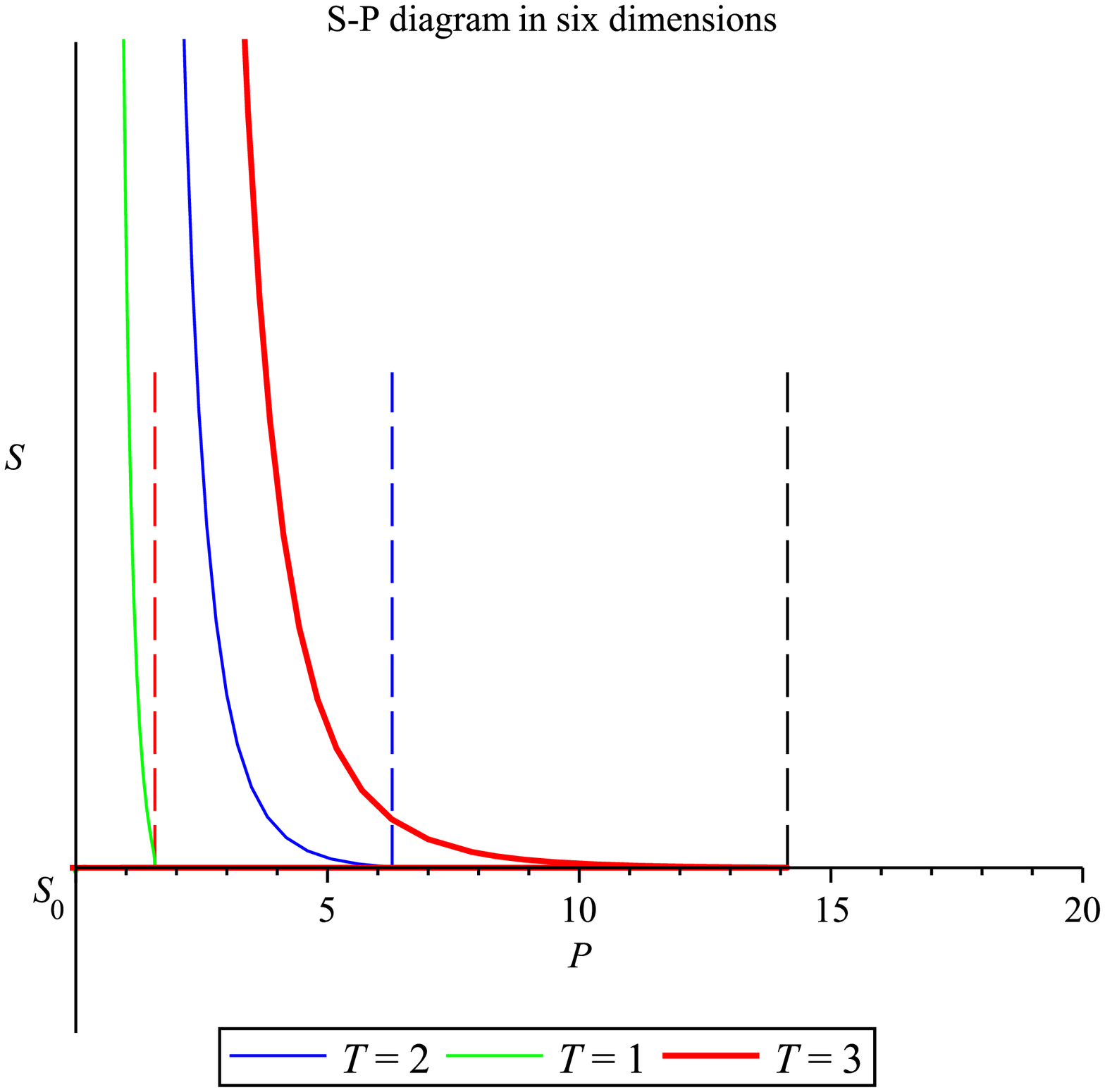}
\includegraphics[width=0.31\textwidth]{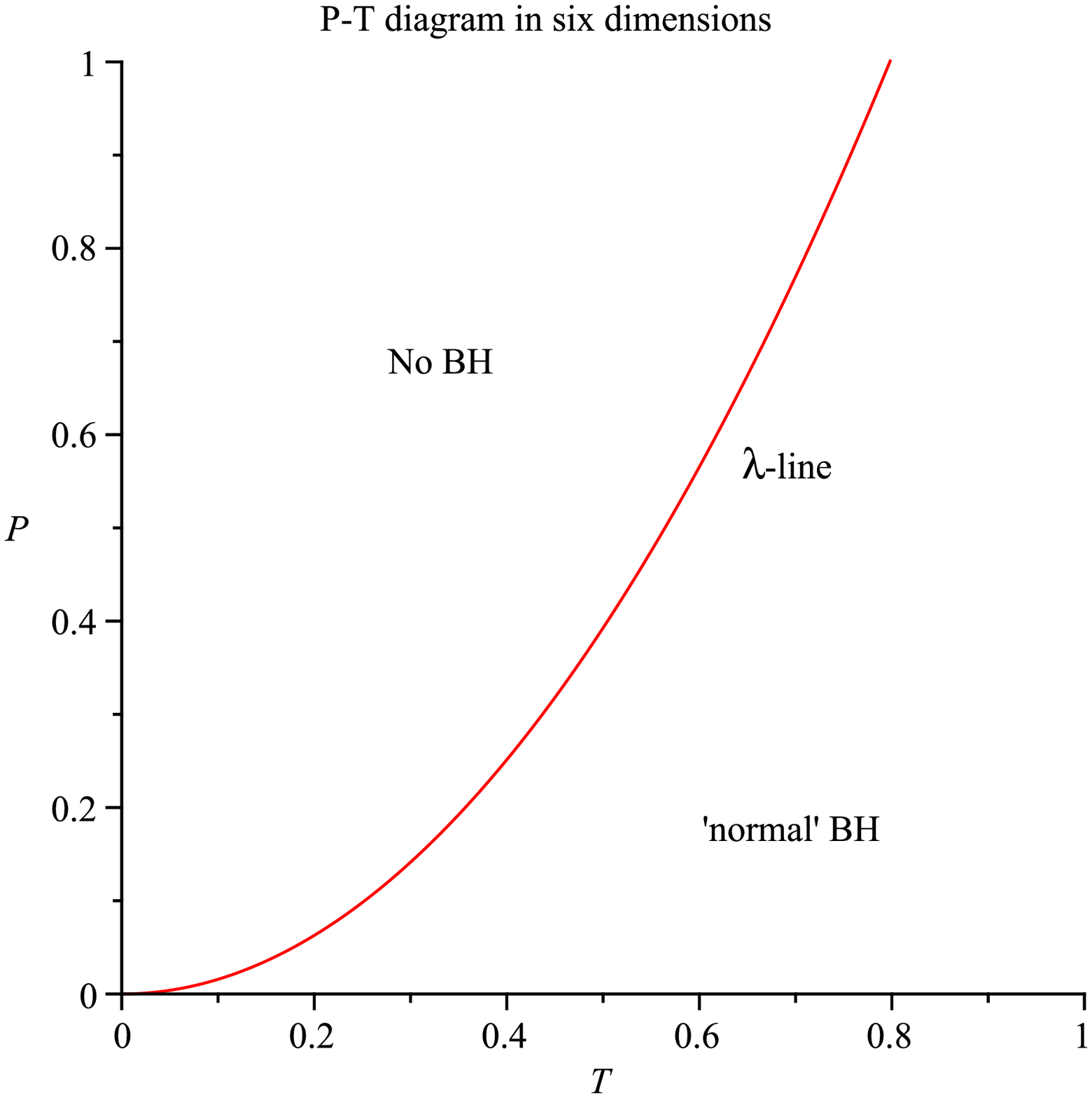}
\caption{
 Curves of $C_P-T$ and $S-P$, and $P-T$ parameter space of black holes with $\epsilon=\epsilon_c$ in six dimensions. The dashed lines highlight the minimal temperature, which could be treated as a physical temperature constraint and cancel the ``superfluid" black hole phase. Therefore there is no $\lambda$ phase transition in six dimensions even the specific heat diverges.
}
\label{6D-2}
\end{center}
\end{figure}

To interpret the strange critical point, we could study the specific heat $C_P$. One can look at the
$C_P-T$ curve in the right plot of Fig.\ref{6D-2}.
The specific heat $C_P$ diverges at the critical point (phase transition point)
\begin{align}\label{Tcri6}
  T=\sqrt{\frac{2P}{\pi}},
\end{align}
while the dashed line
highlighting a lower bound of temperature destroys the $\lambda$ line. It is easy to calculate the
minimal temperature of six dimensional HL AdS spherical black holes with $\epsilon=\epsilon_c$,
for which the temperature is reduced to
\begin{align}
  T(r_+)=\frac{64P^2\pi^2r_+^4+48P\pi\,r_+^2+1}{8\pi\,r_+(1+8P\pi\,r_+^2)}.
\end{align}
Considering the first order derivative
\begin{align}
  T'=\frac{(8P\pi\,r_+^2-1)^3}{8\pi\,r_+^2(1+8P\pi\,r_+^2)^2},
\end{align}
it leads to the minima at $r_+=\frac{1}{\sqrt{8\pi\,P}}$, i.e.
\begin{align}\label{Tmin}
  T_{min}=\sqrt{\frac{2P}{\pi}},
\end{align}
which is exactly the phase transition point (Eq.\ref{Tcri6}). This temperature bound could be treated as
a physical temperature constraint, which cancels the ``superfluid" black hole phase.
This physical temperature constraint is equivalent to a upper bound of pressure, i.e. $P\leq\frac{\pi\,T^2}{2}$, for arbitrary temperature.
Then in the $S-P$ diagram in the middle one of Fig.\ref{6D-2}, one can be also observed that there is no (``superfluid") black holes for $P>\frac{\pi\,T^2}{2}$, which correspond to the exact $\lambda$ line in $P-T$ parameter space as shown in the right plot in Fig.\ref{6D-2}.
Therefore, even six dimensional HL AdS black holes
exhibit infinitely many critical points in $P-\nu$ plane and the divergent points for specific heat, there does not
exist $\lambda$ phase transition, as they
only contain the ``normal" black hole phase and the ``superfluid" black hole phase disappears in the $P-T$ parameter space due to the physical temperature constraint (Eq.\ref{Tmin}).

\section{Critical phenomena in $\epsilon-r_+$ plane}\label{critical exponents}
\label{critical phenomena-1}
Because of the existence of infinite critical points, it is not able to calculate the critical exponents.
Actually, to study the critical exponents for $\lambda$ phase transitions of liquid ${}^4$He,
thermodynamic pressure is no longer the appropriate ordering field \cite{Weichman:2001zz}.
For the ``superfluid" black hole, pressure should be instead of other parameters \cite{Hennigar:2016xwd}.
As for HL AdS black holes, there is only one option for the appropriate ordering field, i.e. parameter $\epsilon$.
Though $\epsilon$ is a dimensionless quantity, it does characterize the critical phenomena for four and five dimensional HL AdS black holes as shown later.

Firstly, we study the critical points in $\epsilon-r_+$ plane,
for which the thermodynamic variables of EOS should be $\epsilon$, $r_+$(i.e. $\nu$), $T$.
Thus, we should firstly re-write the EOS by rearranging the expression for temperature Eq.\ref{temperature} for the chosen ordering field $\epsilon$, which behaviors as
\begin{align}
  \epsilon(T,r_{+})=\pm\sqrt{\frac{(d(d-1)k+32\pi\,P\,r_+^2)}{d(d-1)^2k}\bigg((d-1)-\frac{32\pi\,P\,r_+^2}{8\pi\,r_+T-(d-4)k}\bigg)}.
\end{align}
We follow the conditions
\begin{align}
  \frac{\partial \epsilon}{\partial r_+}=0,\quad\,\frac{\partial^2 \epsilon}{\partial r_+^2}=0,
\end{align}
to find the critical points.
However, the direct differentiation of $\epsilon$ is too complicated, we
also prefer to employ the implicit differentiation on EOS (Eq.\ref{EOS}) and the above equations.
After a careful calculation, we obtain the same two conditions Eq.\ref{ceq1} and Eq.\ref{ceq2}.
Thus, one can easily get the  critical point:
\begin{align}
\epsilon_{c}&=\pm2\sqrt{\frac{(2(d-5)X\pi+2d-7)}{3d(d-4)}},\\
r_c&=\frac{1}{4}\sqrt{\frac{k(d-1)(12\pi\,X+2-d)}{6\pi\,P}},\\
  T_c&=4kX\sqrt{\frac{6\pi\,P}{k(d-1)(12\pi\,X+2-d)}},
\end{align}
where $X$ has the value as Eq.\ref{Xvalue} and $P$ is a positive constant indicating AdS spacetime.

\begin{table}
\begin{center}
\begin{tabular}{|c|c|c|c|}
\hline
Dimensions & $\epsilon_c$&  $r_c$ & $T_c$  \\
\hline
Four  & $\pm\frac{2}{9}\sqrt{9+6\sqrt{3}}$&  $\sqrt{\frac{(2\sqrt{3}-1)}{48\pi\,P}}$ & $\sqrt{\frac{4P}{(2\sqrt{3}-1)\pi}}$\\
\hline
Five & $\pm\frac{2}{3}\sqrt{2}$&  $\sqrt{\frac{1}{8\pi\,P}}$ & $\sqrt{\frac{2P}{\pi}}$ \\
\hline
Six & $\pm\frac{2}{5}\sqrt{5}$&  $\sqrt{\frac{1}{8\pi\,P}}$ & $\sqrt{\frac{2P}{\pi}}$ \\
\hline
\end{tabular}
\caption{The  critical point of HL AdS spherical black holes in $\epsilon-r_+$ plane.}\label{table2}
\end{center}
\end{table}

\begin{figure}[h!]
\begin{center}
\includegraphics[width=0.32\textwidth]{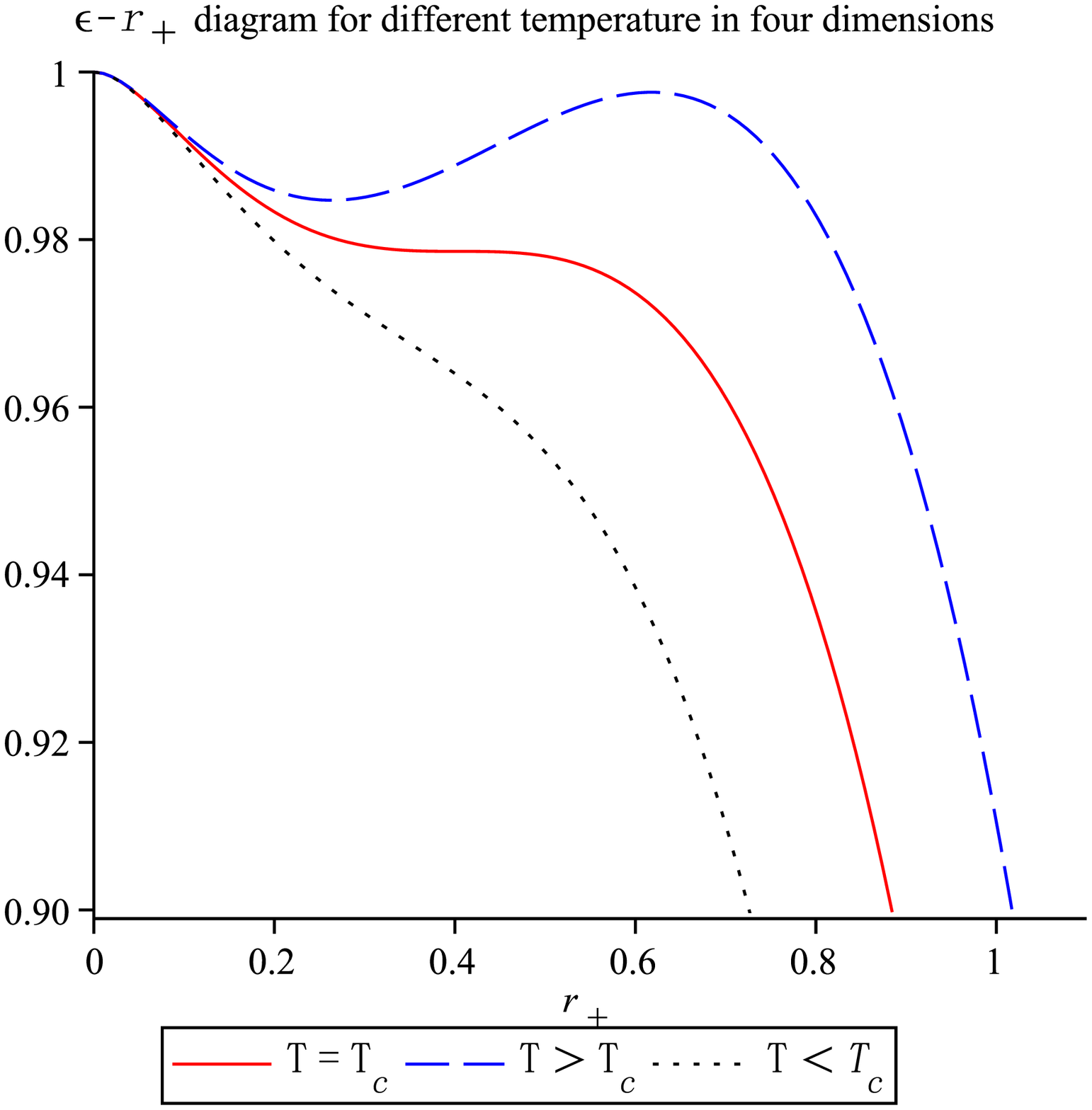}
\includegraphics[width=0.32\textwidth]{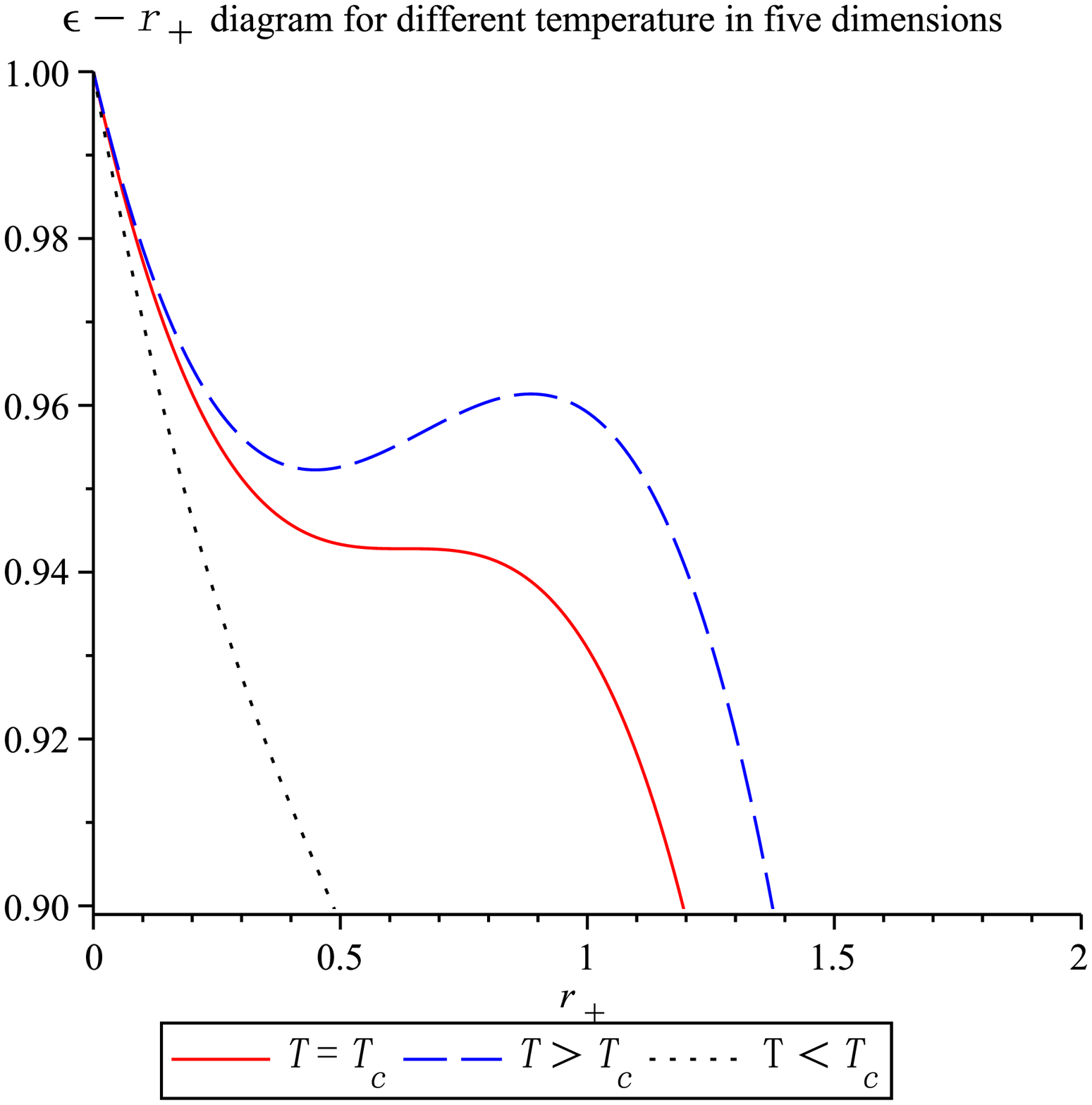}
\includegraphics[width=0.32\textwidth]{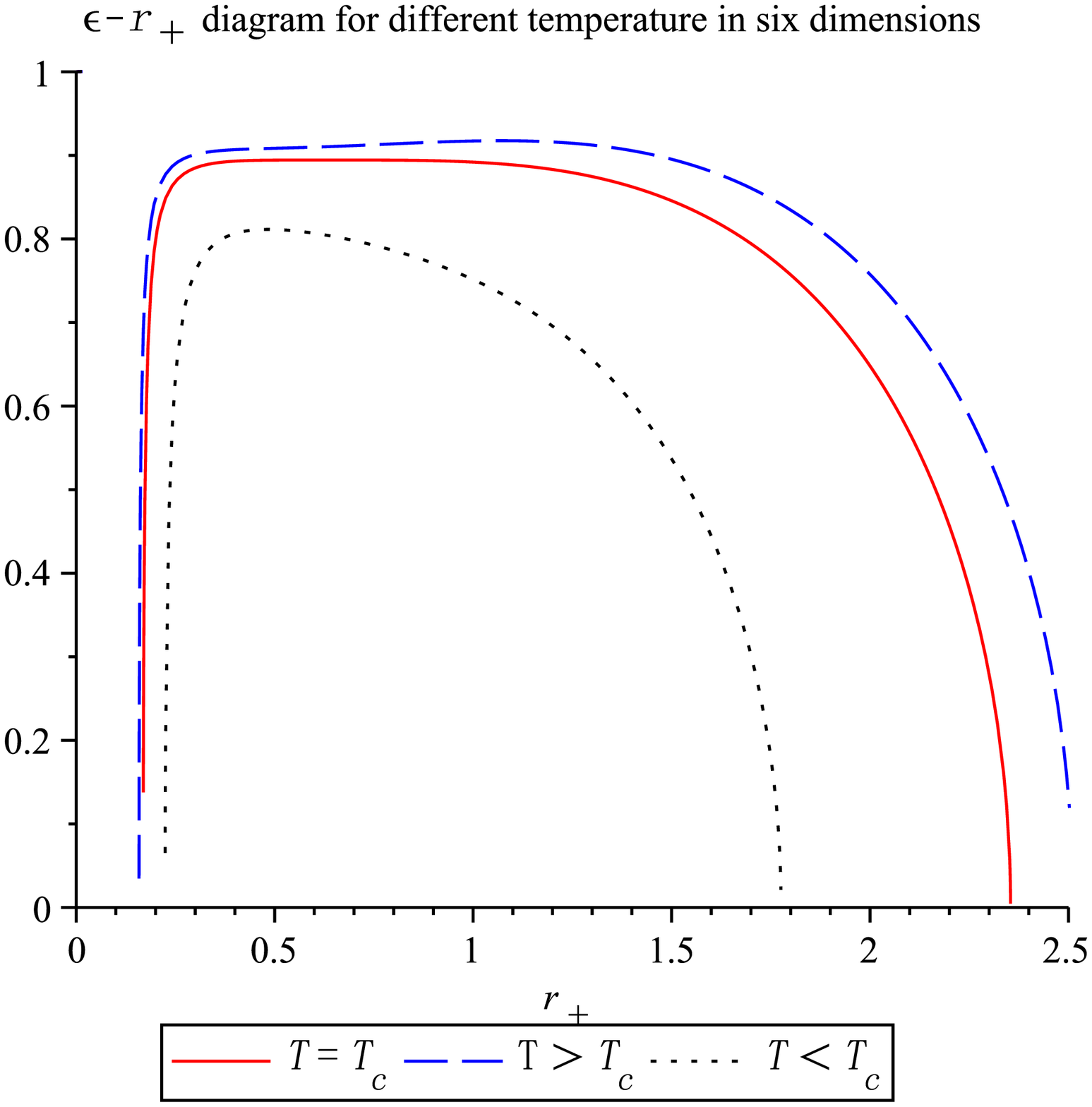}
\caption{Curves of $\epsilon-r_+$ with $P=\frac{1}{10}$ in four, five and six dimensions for different temperature.
In four and five dimensions, one can observe the $\epsilon-r_+$ oscillatory behavior
when $T>T_c$, while no oscillatory behavior for six dimensions.
}
\label{4D5D-A}
\end{center}
\end{figure}

Following the discussions in Section.\ref{PV},
it is shown that only four, five and six dimensional HL AdS black
holes with spherical horizon have  physical critical point, as shown in Table.\ref{table2}.
Besides, in four and five dimensions, there exist the $\epsilon-r_+$ oscillatory behavior when $T>T_c$ as shown in Fig.\ref{4D5D-A}; and the classical ``swallow tail" characterizing the small/large black holes phase transition
when $\epsilon>\epsilon_c$ as shown in $G-T$ diagrams of Fig.\ref{4D5D-1}.
In six dimensions, it is easy to check that the critical point leads to $\frac{\partial^3\epsilon}{\partial\,r_+^3}=0$.
As a result, one can never find the $\epsilon-r_+$ oscillatory behavior as shown in Fig.\ref{4D5D-A}, and no first order phase transition as shown in $G-T$ diagrams of Fig.\ref{6D-3}.

Finally, we consider the critical exponents in four and five dimensions.
After introducing the following dimensionless quantities
\begin{align}
  \Xi=\frac{\epsilon}{\epsilon_c},\quad\,\omega=\frac{r_+}{r_c}-1,\quad\,t=\frac{T}{T_c}-1,
\end{align}
the ordering field can be expanded near the critical point as
\begin{equation}
\Xi=
\left\{ \begin{array}{ll}
1+0.115t+0.234t\omega-0.042\omega^3+\mathcal {O}(t\omega^2,\omega^4),
\quad d=3, \\
1+\frac{1}{4}t+\frac{3}{8}t\omega-\frac{1}{16}\omega^3+\mathcal {O}(t\omega^2,\omega^4), \qquad\qquad\qquad d=4,
\end{array} \right.
\end{equation}
where the coefficients for four dimensional case all have complicated forms, and their approximate value are given.
The above expansions both have the same form as that for the van der Waals fluid and the RN-AdS black hole \cite{Kubiznak:2012wp}.
As the system can be characterized by the following critical exponents:
\begin{align}
  C_{\omega}\propto|t|^{-\alpha},\quad\,\omega\propto|t|^{\beta},
  \quad\,\kappa_t=-\omega^{-1}(\frac{\partial \omega}{\partial \Xi})|_{t}\propto|t|^{-\gamma},\quad\,\Xi\propto|\omega|^{\delta},
\end{align}
one can obtain that
\begin{align}
  \alpha=0,\quad\beta=\frac{1}{2},\quad\gamma=1,\quad\delta=3.
\end{align}
Especially, $\omega\propto\sqrt{t}$ indicates that phase transition appears when $T>T_c$.
Moreover, it is easily to check that they obey the scaling symmetry like the ordinary thermodynamic systems,
in particular coincide with those for a superfluid \cite{Weichman:2001zz}.

\section{Critical phenomena in $P-r_+$ plane with ``temperature" $\epsilon$}
\label{critical phenomena-2}
It is interesting to find that the parameter $\epsilon$ controls the $P-r_+$ oscillatory behavior
instead of the temperature $T$, other than the pressure $P$, as shown in $P-r_+$ diagrams of Fig.\ref{4D5D-1}. This indicates that there is still a critical phenomena in $P-r_+$ plane.
For this case, pressure could still be treated as the ordering field,
while $\epsilon$ should be considered as ``temperature".
This is different from that for the first ``superfluid" black holes \cite{Hennigar:2016xwd}.

To study the corresponding critical phenomena,
we can still follow the procedure studying $P-r_+$ criticality in Section.\ref{PV}, which leads to the same three critical equations: Eq.\ref{EOS}, Eq.\ref{ceq1} and Eq.\ref{ceq2}.
As the thermodynamic variables of EOS for this case should be $P$, $r_+$(i.e. $\nu$), $\epsilon$,
we can find the following  critical point:
\begin{align}
  P_c&=\frac{(d-1)(12\pi\,X+2-d)}{96\pi\,kX^2}T^2,\\
  r_c&=\frac{kX}{T},
\end{align}
with the critical ``temperature"
\begin{align}
  \epsilon_{c}=\pm2\sqrt{\frac{(2(d-5)X\pi+2d-7)}{3d(d-4)}},
\end{align}
where $X$ takes the value as Eq.\ref{Xvalue} and $T$ is a positive constant.

\begin{table}
\begin{center}
\begin{tabular}{|c|c|c|c|}
\hline
Dimensions &  $P_c$ & $r_c$ & $\epsilon_c$ \\
\hline
Four &  $\frac{(2\sqrt{3}-1)\pi\,T^2}{4}$ & $\frac{\sqrt{3}}{6\pi\,T}$ & $\pm\frac{2}{9}\sqrt{9+6\sqrt{3}}$\\
\hline
Five &  $\frac{\pi\,T^2}{2}$ & $\frac{1}{2\pi\,T}$ & $\pm\frac{2}{3}\sqrt{2}$\\
\hline
Six &  $\frac{\pi\,T^2}{2}$ & $\frac{1}{2\pi\,T}$ & $\pm\frac{2}{5}\sqrt{5}$ \\
\hline
\end{tabular}
\caption{The  critical point of HL AdS spherical black holes in $P-r_+$ plane with ``temperature" $\epsilon$.}\label{table3}
\end{center}
\end{table}

\begin{figure}[h!]
\begin{center}
\includegraphics[width=0.32\textwidth]{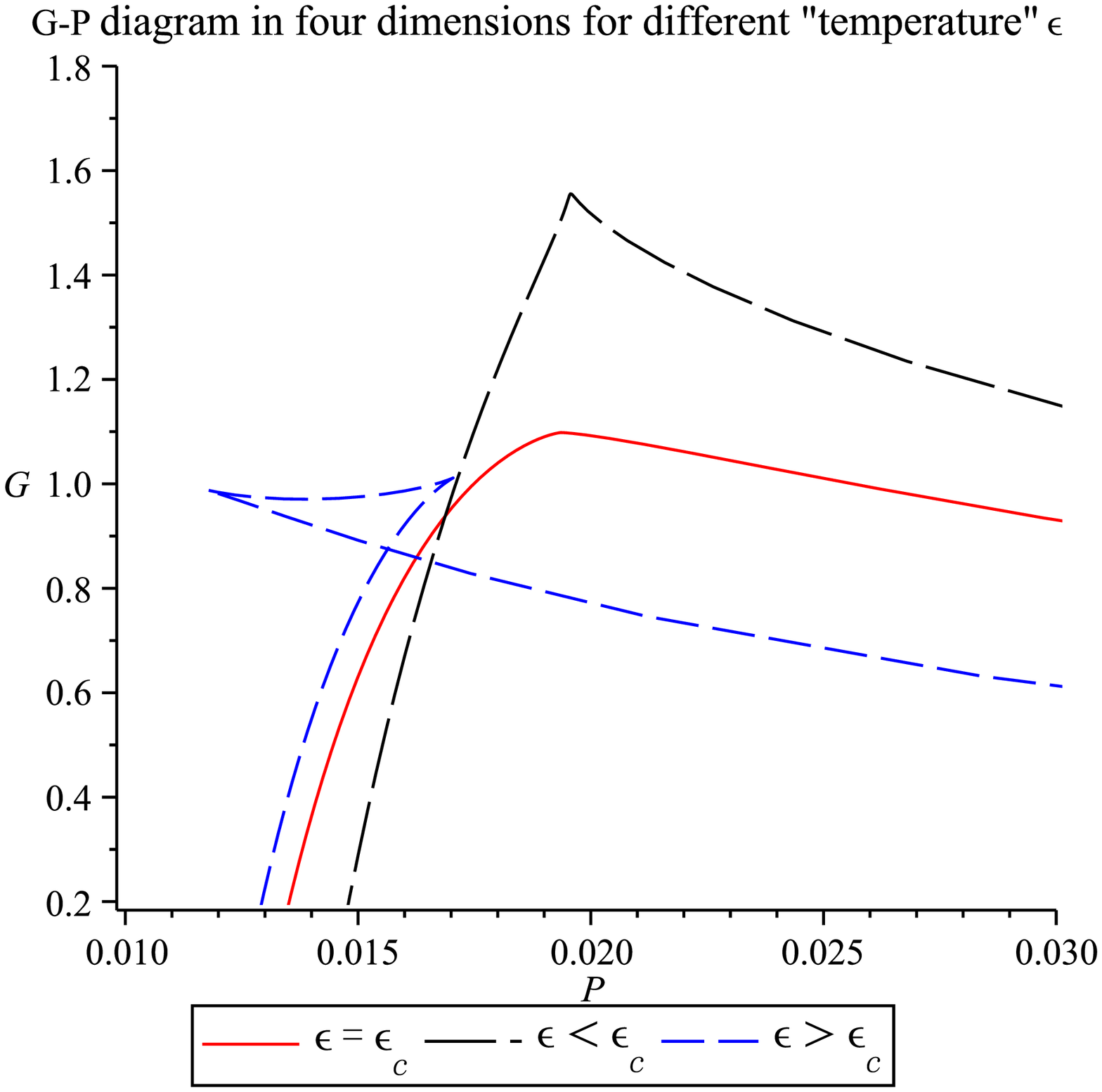}
\includegraphics[width=0.32\textwidth]{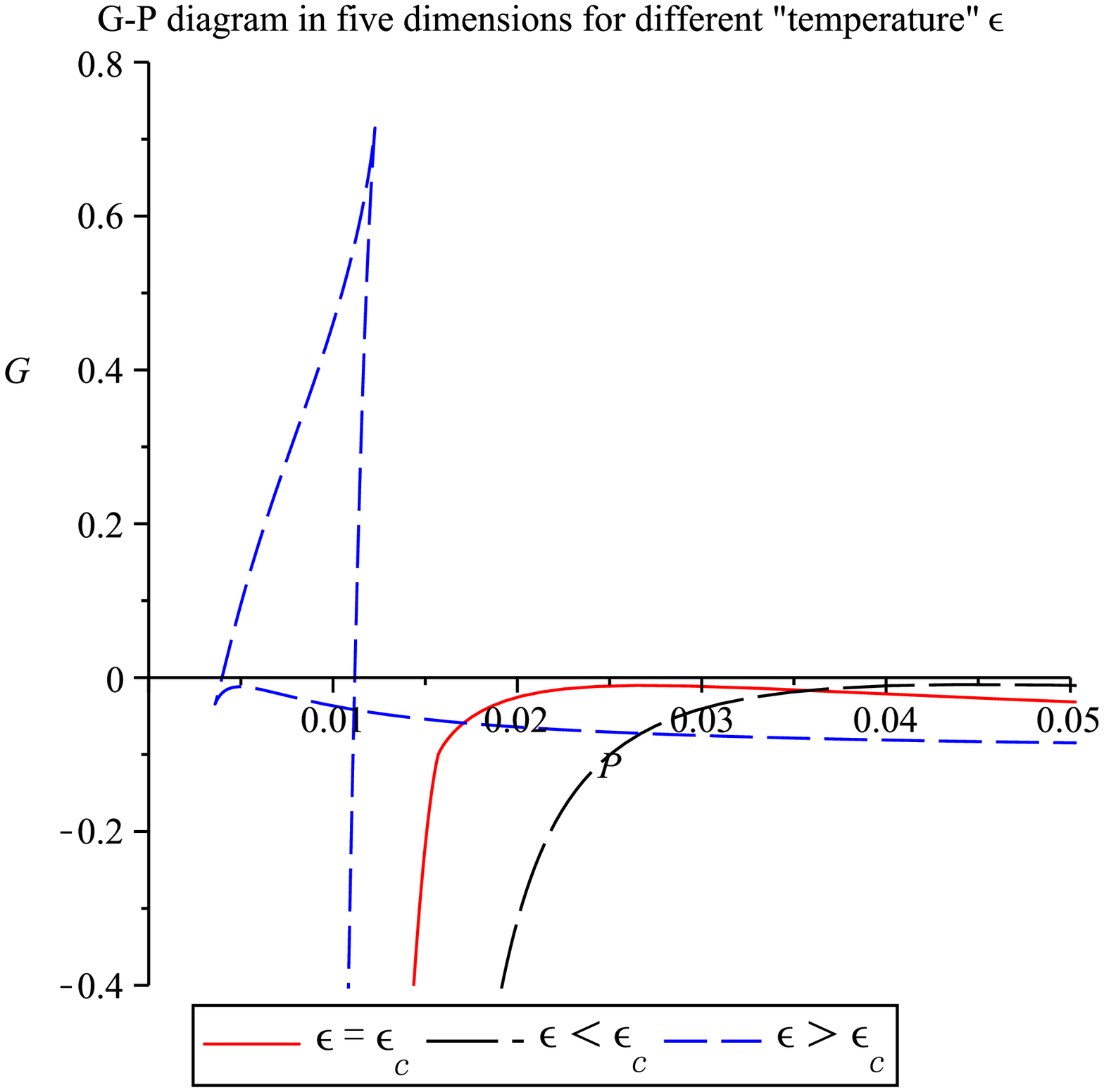}
\includegraphics[width=0.32\textwidth]{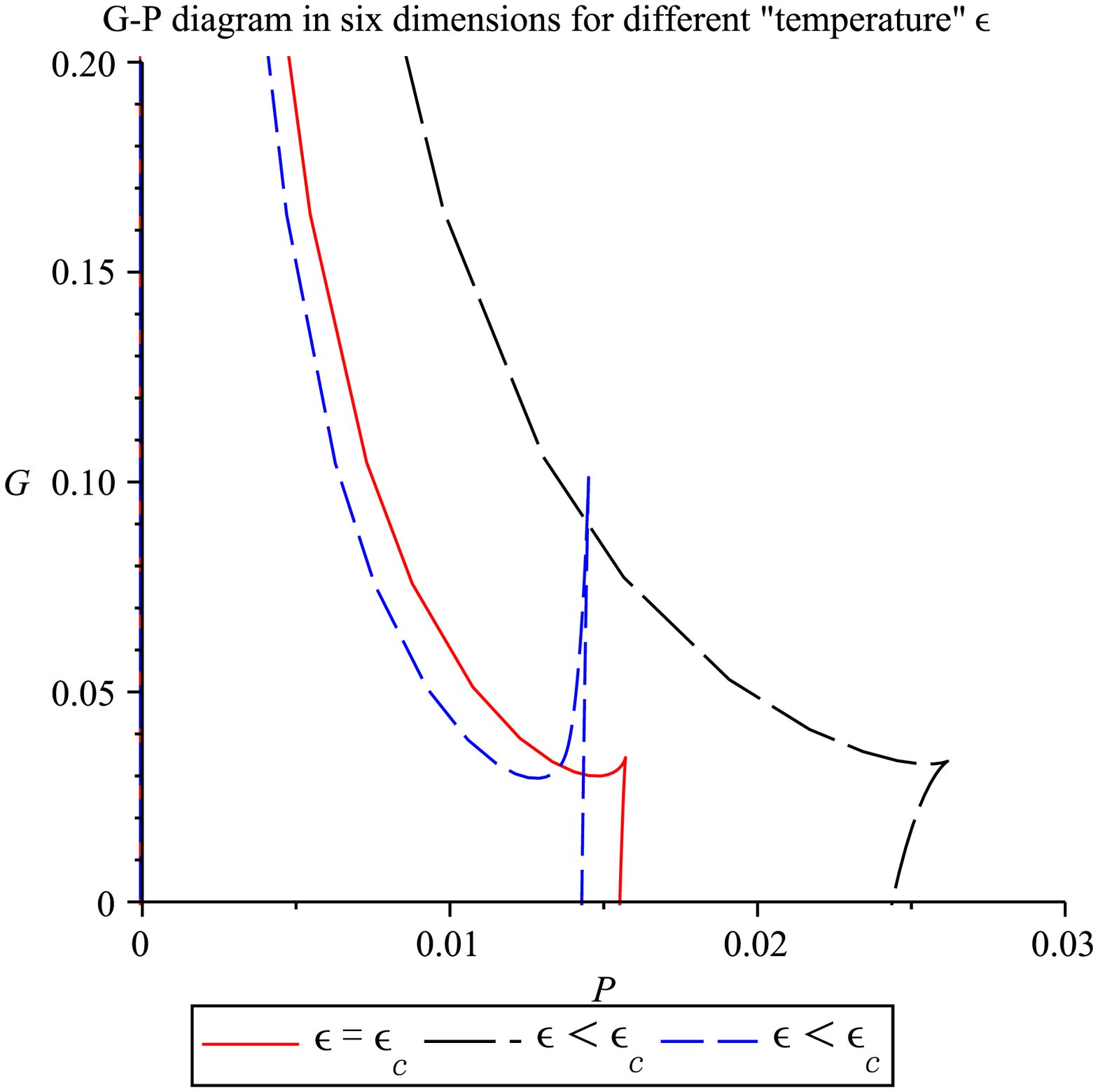}
\caption{Curves of $G-P$ with $T=\frac{1}{10}$ in four, five and six dimensions for different ``temperature" $\epsilon$.
In four and five dimensions, one can observe the classical ``swallow tail" characterizing the small/large black holes phase transition when ``temperature" $\epsilon>\epsilon_c$, while no ``swallow tail" for six dimensions.
}
\label{4D5D-B}
\end{center}
\end{figure}

As similar as the discussions in Section.\ref{PV},
one can find that only four, five and six dimensional HL AdS black
holes with spherical horizon have  physical critical point, as shown in Table.\ref{table3}.
In four and five dimensions, when ``temperature" $\epsilon>\epsilon_c$, there exist the $P-r_+$ oscillatory behavior (Fig.\ref{4D5D-1}), and the classical ``swallow tail" characterizing the small/large black holes phase transition as shown in $G-P$ diagrams of Fig.\ref{4D5D-B}.
As for the case of six dimensions, it is easy to find that the critical point leads to $\frac{\partial^3P}{\partial\,r_+^3}=0$.
Therefore, one can not observe the $P-r_+$ oscillatory behavior as shown in Fig.\ref{6D-3}, and no first order phase transition as shown in Fig.\ref{4D5D-B}.

Then, we can also calculate the critical exponents in four and five dimensions near the critical point.
Similarly, we begin with the following dimensionless quantities
\begin{align}
  p=\frac{P}{P_c},\quad\,\omega=\frac{r_+}{r_c}-1,\quad\,\tau=\frac{\epsilon}{\epsilon_c}-1.
\end{align}
The EOS can be reduced to the dimensionless case, for which its Taylor series expansion at the critical point
takes the following forms
\begin{equation}
p=
\left\{ \begin{array}{ll}
1-\frac{(56\sqrt{3}+94)}{11}\tau+(10\sqrt{3}+18)\tau\omega-(\sqrt{3}-1)\omega^3+\mathcal {O}(\tau\omega^2,\omega^4),
\quad d=3, \\
1-8\tau+12\tau\omega-\frac{1}{2}\omega^3+\mathcal {O}(\tau\omega^2,\omega^4), \qquad\qquad\qquad\qquad\qquad\qquad d=4.
\end{array} \right.
\end{equation}
They still both have the same form as that for the van der Waals fluid and the RN-AdS black hole \cite{Kubiznak:2012wp}.
After introducing the following critical exponents:
\begin{align}
  C_{\omega}\propto|\tau|^{-\alpha},\quad\,\omega\propto|\tau|^{\beta},
  \quad\,\kappa_\tau=-\omega^{-1}(\frac{\partial \omega}{\partial p})|_{\tau}\propto|\tau|^{-\gamma},\quad\,p\propto|\omega|^{\delta},
\end{align}
It is easily to find
\begin{align}
  \alpha=0,\quad\beta=\frac{1}{2},\quad\gamma=1,\quad\delta=3,
\end{align}
which satisfy the scaling laws of the ordinary thermodynamic systems.
Besides, $\omega\propto\sqrt{\tau}$ indicates that phase transition appears when $\epsilon>\epsilon_c$.

\section{Discussion}
In this paper, we study the extended thermodynamics of general dimensional HL AdS black holes and
present another example of ``superfluid" black holes.
It is found that only four and five dimensional HL AdS black holes with spherical horizon have the $\lambda$ phase transition,
which correspond to the phase transition between ``superfluid" black hole and ``normal" black hole.
After considering the behaviour of entropy, the ``superfluid" black hole phase and  ``normal" black hole phase are  distinguished.
Especially, six dimensional HL AdS black holes
exhibit infinitely many critical points in $P-\nu$ plane and the divergent points for specific heat, for which they only contain the ``normal" black hole phase and the ``superfluid" black hole phase disappears due to the physical temperature constraint;
therefore there is no similar phase transition.
In more than six dimensions, there is no $P-V$ critical behavior.
After identifying parameter $\epsilon$ as the ordering field instead of pressure and temperature,
we study the critical phenomena in different planes of thermodynamical phase space. We also obtain the critical exponents in both planes, which are the same with the van der Waals fluid.

The ``superfluid" black hole is firstly reported in Lovelock gravity with conformally coupled scalar field \cite{Hennigar:2016xwd}, which contains at least four free parameters in the action of theory. Comparing with it, the one in Horava gravity has only one free parameter $\epsilon$.
It is interesting to consider the necessary and sufficient conditions
for general ``superfluid" black hole, i.e.
a black hole EOS must satisfy to display the similar $\lambda$ phase transition, which is still un-clear.
Besides, could the ``superfluid" black holes appear in Einstein gravity?
These are all interesting and left to be the future tasks.

\section*{Appendix: Gibbs free energy and capacity}
\label{appendix}
The Gibbs free energy is
\begin{equation}
G=
\left\{ \begin{array}{ll}
&\frac{3k^2(1-\epsilon^2)}{16P\pi\,r_{+}}+2kr_{+}+\frac{16}{3}P\pi\,r_{+}^3-
\frac{(1024P^2\pi^2r_+^4+128kP\pi\,r_+^2-12k^2(1-\epsilon^2))}{16\pi\,r_{+}(32P\pi\,r_+^2+6k(1-\epsilon^2))}\\
&\times\bigg(4\pi\,r_{+}^2(1+\frac{3k(1-\epsilon^2)\ln(r_+)}{8\pi\,r_+^2}) +S_0\bigg),
\qquad\qquad\qquad\qquad d=3; \\
&\frac{k^2dr_{+}^{d-4}(1-\epsilon^2)}{16(d-2)P\pi}+\frac{4kr_{+}^{d-2}}{(d-1)(d-2)}+\frac{64P\pi\,r_{+}^6}{d(d-1)^2(d-2)}\\
&-\frac{1024P^2\pi^2r_+^4+64k(d-1)(d-2)P\pi\,r_+^2+k^2d(d-1)^2(d-4)(1-\epsilon^2)}{8(d-1)\pi\,r_{+}(32P\pi\,r_{+}^2+kd(d-1)(1-\epsilon^2))}\\
&\times\bigg(\frac{16\pi\,r_+^{d-1}}{(d-1)^2(d-2)}(1+\frac{kd(d-1)^2(d-2)(1-\epsilon^2)}{32(d-2)(d-3)P\pi\,r_+^4})+S_0 \bigg), \qquad\qquad d\geq4.
\end{array} \right.
\end{equation}

The specific heat is
\begin{align}
  C_P&=\frac{r_{+}^{d-3}}{2(d-1)(d-2)P}
  \bigg(32P\pi\,r_{+}^2+kd(d-1)(1-\epsilon^2) \bigg)^2\nonumber\\
  &\times\bigg(1024P^2\pi^2r_{+}^4+64k(d-1)(d-2)P\pi\,r_{+}^2+k^2d(d-1)^2(d-4)(1-\epsilon^2) \bigg)\nonumber\\
  &\times\bigg(32768P^3\pi^3r_{+}^6-1024k(d-1)(3d\epsilon^2-d-4)P^2\pi^2r_{+}^4\nonumber\\
  &-32k^2d(d-1)^2(d-8)P\pi\,r_{+}^2(1-\epsilon^2)-k^3d^2(d-1)^3(d-4)(1-\epsilon^2)^2 \bigg)^{-1}.
\end{align}

\section*{Acknowledgements}
We thank the anonymous referees for helpful comments and suggestions.
Wei Xu was supported by the National Natural Science Foundation of China (NSFC)
under Grant No.11505065, No.11374330 and No.91636111, and the Fundamental Research Funds
for the Central Universities, China University of Geosciences (Wuhan).

\providecommand{\href}[2]{#2}\begingroup
\footnotesize\itemsep=0pt
\providecommand{\eprint}[2][]{\href{http://arxiv.org/abs/#2}{arXiv:#2}}

\end{document}